\documentclass[twocolumn,11pt]{aastex631}

\usepackage{graphicx,amsmath}
\usepackage{color}
\usepackage{booktabs}
\usepackage{dutchcal}

\newcommand{\be}{\begin{equation}}
\newcommand{\ee}{\end{equation}}
\newcommand{\bea}{\begin{eqnarray}}
\newcommand{\eea}{\end{eqnarray}}

\usepackage{color}
\usepackage{ifthen}
\newboolean{editorial}
\setboolean{editorial}{true}
\newcommand{\editorial}[2]{\ifthenelse{\boolean{editorial}}{\textcolor{red}{[\textsf{{{#1}}}: }\textcolor{blue}{\textsf{{#2}}}\textcolor{red}{]}}{}}

\shorttitle{Prospect on the detection of lensed multi-messenger signals of BNS mergers}
\shortauthors{Chen, Lu \& Zeng}

\begin{document}

\title{
Prospect for Detection of Strongly Lensed Multi-messenger Signals of Binary Neutron Star Mergers}
\correspondingauthor{Youjun Lu}
%
%
\author[0000-0001-7952-7945]{Zhiwei Chen}
\affiliation{National Astronomical Observatories, Chinese Academy of Sciences, 20A Datun Road, Beijing 100101, China}
\affiliation{School of Astronomy and Space Sciences, University of Chinese Academy of Sciences, 19A Yuquan Road, Beijing 100049, China}
\author[0000-0002-1310-4664]{Youjun Lu}\thanks{luyj@nao.cas.cn}
\affiliation{National Astronomical Observatories, Chinese Academy of Sciences, 20A Datun Road, Beijing 100101, China}
\affiliation{School of Astronomy and Space Sciences, University of Chinese Academy of Sciences, 19A Yuquan Road, Beijing 100049, China}
\author{Changwen Zeng}
\affiliation{National Astronomical Observatories, Chinese Academy of Sciences, 20A Datun Road, Beijing 100101, China}
\affiliation{School of Astronomy and Space Sciences, University of Chinese Academy of Sciences, 19A Yuquan Road, Beijing 100049, China}

\begin{abstract}

The gravitational lensing of multi-messenger signals from binary neutron star mergers (BNSs), including gravitational waves (GWs), short Gamma-Ray bursts (sGRBs), kilonovae, and afterglows, can serve as a unique probe to constrain the mass of the graviton and cosmological parameters.
In this paper, we estimate the detection rates of lensed electromagnetic counterparts associated with lensed BNS GW events detected by Cosmic Explorer and Einstein Telescope. {For kilonovae and afterglows, we further consider a complementary pointed follow-up strategy targeting pre-identified galaxy-scale lens candidates within the GW localization region.}
By utilizing both numerical and observational constraints on BNS mergers, we find that:
(1) {Future $\gamma$-ray telescopes, even with a sensitivity more than ten times better than that of Fermi-GBM, may only detect lensed sGRB prompt emission at a rate $\sim 0.1$\,yr$^{-1}$, corresponding to $\sim 2\times 10^{-3}$ of detectable lensed BNS GW events.}
(2) {For the known-lens pointed strategy, the identifiable lensed-host fraction is approximately $0.15-0.30$ for the fiducial deep lens-catalog case considered, suggesting a possible gain in per-lens sensitivity for faint kilonovae and afterglows.}
(3) {An RST-like near-infrared facility could detect lensed kilonovae at rates of {approximately $\sim 0.45^{+0.81}_{-0.34}$, $0.55^{+0.98}_{-0.41}$, and $0.078^{+0.139}_{-0.059}$\,yr$^{-1}$ }in the F106, F158, and F213 bands, respectively.}
(4) {Lensed afterglows remain difficult to detect in the optical and radio bands, while ATHENA-like X-ray observations {may detect $0.5-5$} events over ten years.}

\end{abstract}



\section{Introduction}
\label{sec:intro}

Mergers of binary neutron stars (BNSs) are perhaps the most important gravitational wave (GW) sources in the Universe in terms of its diverse potential multi-messenger signals produced in the baryon-rich environment. The subsequent observations of the first BNS merger GW170817 detected by the Laser Interferometer Gravitational Wave Observatory (LIGO) and Virgo, including the short Gamma-Ray burst (sGRB), kilonova, afterglow in the X-ray, optical, and radio bands, open a new window for multi-messenger studies \citep[e.g., ][]{2017ApJ...848L..12A, 2017ApJ...848L..13A, 2017Sci...358.1556C, 2017Natur.551...67P, 2018ApJ...858L..15D, 2019MNRAS.489L..91C, 2019PhRvX...9a1001A}. For example, several significant physical processes may be constrained by such observations, for example, the tidal deformation of neutron stars (NSs) \citep[e.g., ][]{2018PhRvL.120z1103M, 2018PhRvL.121i1102D}, the jet propagation in the ambient medium \citep[e.g.,][]{2018MNRAS.479..588G, 2018ApJ...863...58X}, and even the equation of state (EOS) of NSs \citep[e.g., ][]{2018ApJ...852L..29R, 2018MNRAS.480.3871C, 2018PhRvC..98c5804M}. 

With powerful third-generation GW detectors such as the Cosmic Explorer \citep[hereafter CE,][]{2019BAAS...51g..35R} and the Einstein Telescope \citep[hereafter ET,][]{Hild_2011}, it is anticipated to detect $10^4-10^5$ BNS mergers per year, among which a small fraction $\sim 10^{-3}$ may be strongly gravitational lensed by intervening galaxies or clusters \citep[e.g., ][]{2022MNRAS.509.3772Y, 2023MNRAS.518.6183M, 2023MNRAS.520..702S}. Due to the magnification and time-delay effects, these lensed BNS events with electromagnetic (EM) counterparts are of great significance in many aspects {\citep[e.g.,][]{2025RSPTA.38340134S,2025RSPTA.38340127M,2025RSPTA.38340122L}}. First, lensed BNS mergers with known redshift can be taken as independent probes to constrain the Hubble constant and other cosmological parameters by the ``time delay'' distance since the time delay can be measured accurately, due to the high sampling rate and time resolution of GW detections {\citep[e.g.,][]{2017NatCo...8.1148L, 2020MNRAS.498.3395H,2025RSPTA.38340130B}}. Second, one may also test the speed of GW and constrain the mass of the graviton by comparing the propagation paths of the lensed GW and EM signals \citep[e.g.,][]{2017PhRvL.118i1102F, 2020PhRvD.101j3509M}. Third, due to the time delays between multiple images of lensed BNS mergers, the first image may provide an early warning of the subsequent images, enabling an effective search for their EM counterparts \citep[e.g.,][]{2023MNRAS.518.6183M}. Therefore, it is important to estimate the detection rate of lensed BNSs with multimessenger signals. 

The detection rate of the kilonovae and optical afterglows of lensed GW BNS mergers has been studied via Monte Carlo simulations in a previous work \citep[hereafter MLG+23]{2023MNRAS.518.6183M}, in which it was found that the Roman Space Telescope (RST) and the James Webb Space Telescope (JWST) \citep[]{2006SSRv..123..485G,2013arXiv1305.5422S} have the capability to detect about a few or more such events per year 
{
after the identification of the associated lensed host galaxies,
}
and the EM images of most lensed BNS mergers can be spatially resolved. The results in MLG+23 are obtained by assuming that the intrinsic parameters of all the BNS mergers are chosen as the same as those of GW170817, for example, the ejecta mass $m_{\rm ej}$ and the normalized jet kinetic energy $E^{\rm K}_0$. 
{
However, in the real universe, the component masses $m_1$ and $m_2$ of BNS merger systems can differ from those of GW170817, resulting in variations in the ejecta mass $m_{\rm ej}$ and the initial kinetic energy $E^{\rm K}_0$. Given our current understanding of the physical processes governing BNS mergers, GW170817 likely ranks among the most luminous kilonovae from BNS mergers with diverse physical parameters, despite its relatively small luminosity distance.
} 

We extend MLG+23 by adopting an enhanced unified framework similar to \citet{2022ApJ...937...79C}. This framework directly relates the isotropic energy of sGRB, the light curves (LCs) of kilonova and afterglows in the optical, radio, and X-ray bands, and the images of host galaxies to mock BNS merger samples generated by our binary population synthesis model. {This approach offers a novel perspective on the joint detection of strongly lensed multi-messenger signals from BNS mergers using future telescopes and next generation GW detectors like CE and ET, by emphasizing the importance of the identification of lensed host galaxies. Such advancements will be a nice addition to the search for lensed multi-messenger signals beyond the well-known Target of opportunity (ToO) strategy \citep[e.g.,][]{2024arXiv241104793A,2025RSPTA.38340118R} for the upcoming era of third-generation GW detection. } 

This paper is organized as follows. In Section~\ref{sec:method}, we introduce the main methodology, including the generation of strongly lensed signals of GW events (Section~\ref{sec:gw}) and their associated kilonova (Section~\ref{sec:kilonova}), sGRB (Section~\ref{sec:grb}), afterglow (Subsection~\ref{sec:afterglow}), and host galaxies (Section~\ref{sec:host}) in multibands. In Sections~\ref{sec:gw+sgrb} and \ref{sec:gw+multi}, we present our main results on the joint detection rate of such lensed multi-messenger signals. The discussion and conclusions are given in Section~\ref{sec:con}. Throughout the paper, we adopt the cosmological parameters as $(h_0,\Omega_{\rm m},\Omega_\Lambda)=(0.68,0.31,0.69)$ \citep{Aghanim2020}.
  
\section{Methodology }
\label{sec:method}
{The multi-messenger signals of BNS mergers may be strongly gravitationally lensed by intervening objects, such as galaxies, galaxy clusters and galaxy groups. As for next generation GW detectors such as CE and ET with enhanced sensitivities, the detection horizon for BNS mergers may be enlarged to $z_{\rm H}\sim 1-2$. According to the discussion in Section  4(a) in \citet{2025RSPTA.38340134S}, for $z_{\rm H}>0.5$, the multiple-imaged detections will be dominated by galaxy-scale lenses rather than clusters and groups, especially for those with magnification factor $\mu\lesssim 10$.
Therefore, in this work, we focus on the detectability of multi-messenger signals of BNS mergers lensed by intervening galaxies with CE and ET. }
{The galaxy strong lensing is mainly dominated by the elliptical early-type galaxies, which is often modeled as singular isothermal ellipsoid (SIE) in literature \citep[e.g.,][]{2018MNRAS.476.2220L, 2023MNRAS.518.6183M}. Here we note that the late type galaxies may also contribute to the lensing, which could lead to an underestimation of our results, since the first two lensed supernovae discovered in synoptic optical surveys were not lensed by a massive early-type galaxy but rather by late-type galaxy \citep[e.g.,][]{2014Sci...344..396Q,2017Sci...356..291G}. 
}

{To account for more realistic and complex configurations including non-isothermal image pairs, we adopt an elliptical power-law density profile (EPL) \citep{2015A&A...580A..79T} supplemented by external shear to model the lensing galaxies, following \citet{2021ApJ...921..154W}.}
The EPL model is characterized by the spectral index of the density profile $\lambda$, the velocity dispersion $\sigma_{\rm v}$, and the axis ratio $q$. 
In this work, we adopt a Gaussian distribution for the density-profile index $\lambda$, with a mean of $2$ and a scatter of $0.2$, consistent with measurements from the SLACS survey of early-type strong-lens galaxies \citep[e.g.,][]{2009ApJ...703L..51K}. Notably, when $\lambda=2$, the EPL profile reduces to the SIE model. 

As for the velocity dispersion,  we assume a Schechter form for the elliptical galaxies and hot components of spiral galaxies, i.e., \citep[see][]{2007ApJ...658..884C, Piorkowska:2013eww} 
\begin{equation}
\frac{dn(\sigma_{\rm v},z_{l})}{d\ln \sigma_{\rm v}}=n_z \frac{\beta}{\Gamma(\alpha/\beta)}
\left(\frac{\sigma_{\rm v}}{\sigma_z}\right)^{\alpha}\exp{\left[-\left(\frac{\sigma_{\rm v}}{\sigma_z}\right)^{\beta}\right]},
\label{eq:lens}
\end{equation}
and
\begin{equation}
{n_z = n_{0}(1+z)^{\kappa_{n}};\quad \sigma_z = \sigma_{\rm v0}(1+z)^{\kappa_{\rm v}}}.
\end{equation}
Here $\sigma_{\rm v0}$ is the characteristic velocity dispersion, $\alpha$ is the power law index at the low-velocity end, $\beta$ is the high-velocity exponential truncated index, and $\Gamma(\alpha/\beta)$ is the Gamma function. The reference parameters in this formula are chosen to be the canonical values, i.e., $(n_0, \sigma_{\rm v0}, \alpha, \beta) = (0.008h^{3} {\rm Mpc}^{-3}, 161{\rm km \, s^{-1}}, 2.32, 2.67)$ \citep{2007ApJ...658..884C,Piorkowska:2013eww}. For the evolution parameters $\kappa_n$ and $\kappa_{\rm v}$, we adopt the fitting results from \citet{2021MNRAS.503.1319G}, i.e., $\kappa_n = -1.18$ and $\kappa_{\rm v}=0.18$. In addition, the distribution of the axis ratio $P(q)$ is assumed to be a truncated Gaussian distribution in the range of $[0.2,1]$ with a mean of $0.7$ and a standard deviation of $0.16$ to match the observations on early-type galaxies \citep{2003ApJ...594..225S}. In addition, both the external shear $\gamma_1$ and $\gamma_2$ are drawn from a normal distribution centered at $0$ with a width of $0.5$ \citep[e.g.,][]{2015ApJ...811...20C}.  

{
In this work, we generate our mock BNS merger population using 
the $\boldsymbol{\alpha10.\rm kb\beta0.9}$ model from \citet{2022MNRAS.509.1557C}, by incorporating the observational extinction-corrected SFR from \citet{annurev:/content/journals/10.1146/annurev-astro-081811-125615} and the mean metallicity redshift evolution obtained in \citet{2016Natur.534..512B}. Several key physical processes are considered in this model, including common envelope evolution, natal kicks, and mass ejection during the secondary SN. When implemented within the cosmological simulations of galaxy formation and evolution and/or Milky Way-like galaxies, \citet{2022MNRAS.509.1557C} found that the model reproduces both Galactic BNS observations and local BNS merger rates inferred from GW data via Bayesian method. According to this model, the number distribution of the BNS mergers can be estimated as \citep[e.g.,][]{2021MNRAS.500.1421Z, 2023ApJ...953...36C}
\begin{equation}
\frac{d^3\dot{N}}{d{m_1}dq dz}=\frac{{R}(z,{m_1},q)}{1+z} \frac{dV(z)}{dz},  
\label{source}
\end{equation}
where $m_1$ is the primary mass, $q$ is the mass ratio, and $R(z,m_1,q)$ is the merger rate density taken from this model with the primary mass in the range from $m_1$ to $m_1+ dm_1$, the mass ratio in the range from $q$ to $q+dq$ at redshift $z$. Here we rescale $R(z,m_1,q)$ {by the observation of local merger rate density from LIGO-Virgo-KAGRA collaboration \citep{2025arXiv250818083T}, i.e., $R_0\sim 89_{-67}^{+159} \rm Gpc^{-3} yr^{-1}$.}
}

{
According to Equation~\eqref{source}, we use Gibbs sampling to generate $10^7$ mock BNS mergers with component masses $m_1$ and $m_2$ over the redshift range $z\in [0,5)$. For each system, we compute NS radii $r_1$ and $r_2$ assuming the {SLy} equation of state, whose non-rotating maximum NS mass is $M_{\rm TOV}=2.06 M_{\odot}$ \citep[e.g., ][]{2001A&A...380..151D}. This yields $10^7$ parameter tuples  $(z,m_1,m_2,r_1,r_2)$, from which we directly infer the ejecta mass $m_{\rm ej}$ and remnant disk mass $m_{\rm disk}$, key parameters for predicting the kilonova magnitudes. For the sGRB and multiband afterglow signals, we estimate the isotropic-equivalent energy under the Blandford-Znajek jet-launching mechanism \citep[e.g.][]{1977MNRAS.179..433B}. Further details are provided in Appendix ~\ref{app:B1}.
}

{
We also note that sources aligned suitably with the lensing axis can produce multiple images. In this work, we classify events as double, triple, or quadruple image cases (neglecting the extremely rare quintuple-image case), whenever there are at least $2$, $3$, or $4$ GW images detected and an equal number of corresponding lensed EM counterparts, according to the detectability criteria introduced below.
}

\subsection{Lensed GWs}
\label{sec:gw}

We generate the GW waveform for each BNS merger via the standard package pyCBC \citep[][]{2019PASP..131b4503B} by adopting the phenomenological model {IMRPhenomPv2NRTidalv2}. The total strain $h(f)$ received by a GW detector is
\begin{equation}
h(f)=F_+ (f)h_+(f)+F_{\times}(f)h_{\times}(f),
\end{equation}
where $F_+$ and $F_{\times}$ are the pattern functions of the detector. 
We define the whitened GW data sets of a GW network composed of $n$ detectors ($n=1$ for a single detector) as
\begin{equation}
\hat{\mathbf{d}}(f)=\left(\frac{A_1(f)h_1(f)}{\sqrt{S_{1}(f)}},\frac{A_2(f)h_2(f)}{\sqrt{S_{2}(f)}}, ..., \frac{A_{n}(f)h_{n}(f)}{\sqrt{S_{n}(f)}}\right),
\end{equation}
where $A_{n}=e^{-2\pi i f((\hat{r}_{n}-\hat{r}_{1})\cdot \hat{n}_{\rm GW})}$ is the phase transfer function, $\hat{r}_{n}$ is the location of the $n$-th detector, $\hat{n}_{\rm GW}$ is the unit direction vector of the GW source, and $S_{n}$ denotes the one-sided power spectrum of the corresponding $n$-th GW detector \citep[][]{2010PhRvD..81h2001W}. 

The optimal squared signal-to-noise ratio (S/N) is then given by
\begin{equation}
\varrho_{GW}^2=\left\langle\hat{\mathbf{d}}(f)|\hat{\mathbf{d}}(f)\right\rangle,
\label{eq:SNR}
\end{equation}
where the angular bracket denotes an inner product. For any two vector functions $\hat{\mathbf{a}}(f)$ and $\hat{\mathbf{b}}(f)$, this inner product is defined as
\begin{equation}
\langle \hat{\mathbf{a}}(f)| \hat{\mathbf{b}}(f)\rangle=2\sum_{j} \int_{f_{\text {min }}}^{f_{\text {max }}}\left\{a_j(f) b_j^{*}(f)+a_j^{*}(f) b_j(f)\right\} d f ,
\end{equation}
where $j$ denote for the $j$-th component of the vector, ${f_{\rm min}}$ and ${f_{\rm max}}$ are the lower and upper frequency limits of the GW waveforms. Once a BNS merger have multiple images and the S/N ($\varrho_{\rm GW}$) of each image is larger than $8$, i.e., $\sqrt{\min(\{|\mu_{\rm m}|\})}\varrho_{\rm GW}>8$, where $|\mu_{\rm m}|$ represents a list of the magnification factors for all images, we identify it as a lensed BNS GW event. 
In this paper, we focus on the performance of third-generation GW detectors and their networks, including ET\footnote{ET-D design \citep[][]{Hild_2011} \href{http://www.et-gw.eu/}{http://www.et-gw.eu/} } and CE\footnote{Stage-2 phase \citep[][]{2019BAAS...51g..35R} \href{https://cosmicexplorer.org/}{https://cosmicexplorer.org/}}.

{Note that the above detection criterion is comparably idealized and simplified. In real observation runs, there are several factors that may affect the detection of lensed BNS GW events. First, the duty cycle of GW detectors significantly impacts the detectability of multi-image events. For a lensed system with multiple images separated by time delays ranging from minutes to months, the detectors must be operational at each arrival time to identify the event as a lensed pair. Assuming a projected duty cycle of $\sim 85\%$, and assuming uncorrelated downtimes for long time delays, the joint detection efficiency for a double-image event would scale approximately as $\sim 0.85^2 \approx 0.72$, indicating that the detection rate estimated shall reduce by a factor of $\sim 20-30\%$ due to detector downtime. Second, identifying bona fide lensed pairs from the vast population of unlensed BNS mergers presents a challenge regarding false positives. In practice, matching requires examining the consistency of inferred parameters (e.g., chirp mass, mass ratio, spin, and sky localization) across different triggers \citep[e.g.,][]{2018arXiv180707062H}. While the high merger rate increases the probability of random coincidences, the high SNR expected for next generation detectors allows for extremely precise parameter estimation \citep[e.g.,][]{2023PhRvD.107f3023C}. This precision significantly reduces the overlap of posterior distributions for unrelated events, thereby suppressing false positives.}

\subsection{Lensed Kilonova}
\label{sec:kilonova}

{
We model the light curves of each kilonova using an anisotropic multi-component model \citep[e.g.,][]{2021MNRAS.505.1661B}, including dynamical, viscous and wind ejecta. For the mock BNS population, we infer the ejecta masses $m_{\rm ej}=(m_{\rm dyn}, m_{\rm vis}, m_{\rm wind})$ from the binary component masses via calibrated fits to numerical simulations \citep[e.g., ][]{2017CQGra..34j5014D,2020PhRvD.101j3002K}, to accounting for the diversity of kilonova signals. Detailed method is provided in Appendix ~\ref{app:B1}. 
} 
The other intrinsic parameters, such as opacity, can be fitted by the LCs of AT2017gfo in the u-, g-, r-, i- and z-bands observed by DECAM \citep[e.g.,][]{ 2017ApJ...851L..21V}. In this work, we assume that the values of the following parameters remain the same for all mock BNS mergers, i.e., the energy normalization of the r-process $\epsilon_0=183.4\times10^{18} \rm erg ~g^{-1}s^{-1}$, the lanthanide-rich flat temperature $T_{f}^{\rm LA}=247.8 \rm ~K$, the lanthanide-free flat temperature $T_{f}^{\rm Ni}=4475.4 \rm ~ K$, low-elevation opacity $\kappa_{\rm low}=44.7 \rm ~ cm^2 g^{-1}$, high-elevation opacity $\kappa_{\rm low}=0.43 \rm ~ cm^2 g^{-1}$, wind opacity $\kappa_{\rm wind}=33.5 \rm ~ cm^2 g^{-1}   $ and viscous opacity $\kappa_{\rm vis}=47.8 \rm ~ cm^2 g^{-1}$. Here we also note that the LCs of a kilonova viewing at different $\theta_{\rm v}$ may be somewhat different. Thus, we should consider the viewing angle distribution for BNS mergers obtained by GW detection, which is \citep[e.g., ][]{2011CQGra..28l5023S, 2012ApJ...746...48M}
\begin{equation}
P(\theta_{\rm v})=0.076(1+6\cos^2{\theta_{\rm v}}+\cos^4{\theta_{\rm v}})^{3/2}\sin{\theta_{\rm v}},
\end{equation}
based on the dependence of the sensitivity of the GW detectors on the viewing angle (i.e., the pattern functions of the detectors). 

\begin{figure}
\centering
\includegraphics[width=1.0\columnwidth]{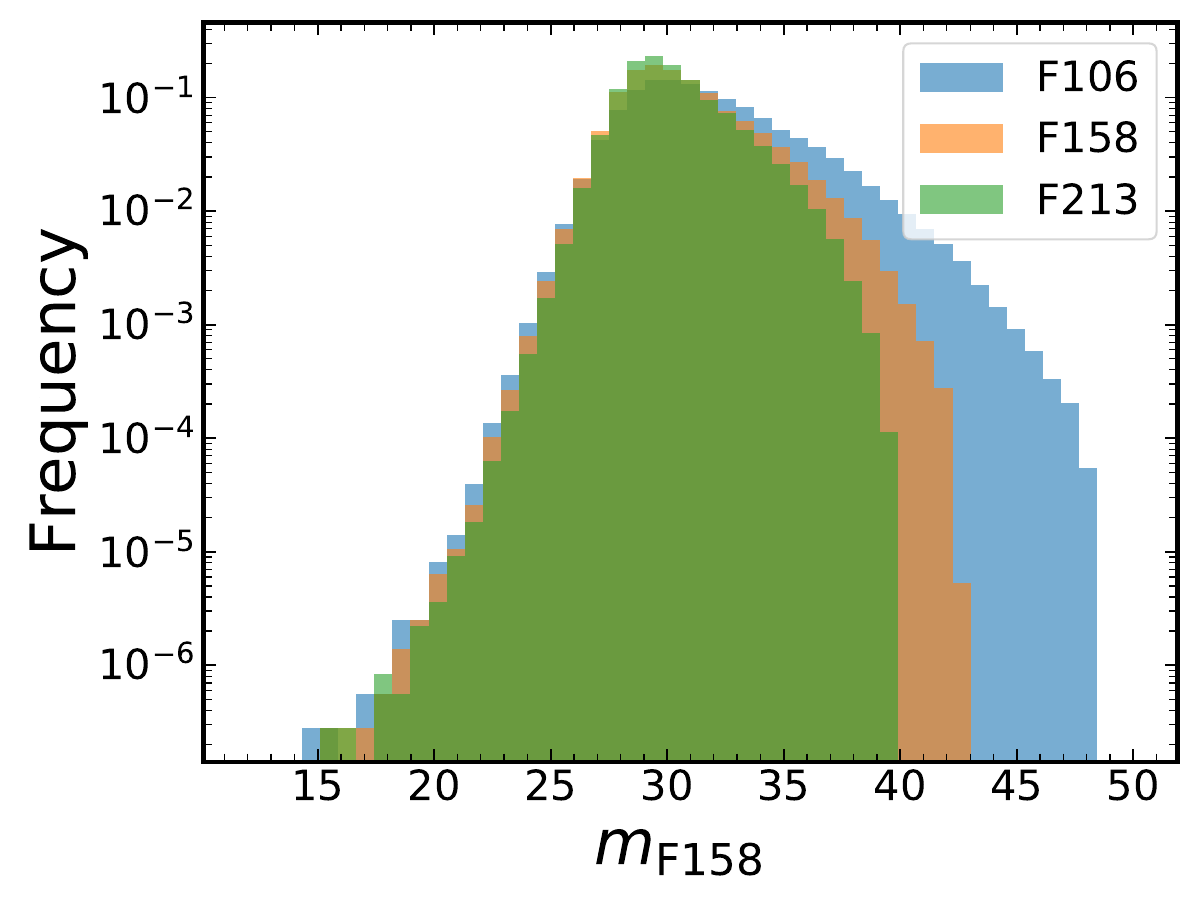}
\caption{ The normalized probability distribution of the apparent magnitudes for the kilonova signals from {the mock BNS mergers, } calculated one day post-merger {over the redshift range of $[0,5)$}, if observed by F106 (blue), F158 (orange), and F213 (green) bands of RST, respectively. }
\label{fig:kilo}
\end{figure}

Figure~\ref{fig:kilo} shows the distribution of the apparent magnitudes for the kilonova signals from the mock BNS mergers 
{
over the redshift range of $[0,5)$. The apparent magnitude of the signal in each band is calculated at the time of one day post-merger.
} 
The blue, orange, and green histograms represent the results if observed by the F106, F158, and F213 bands of RST, respectively. As seen from this figure, the peak magnitude of the distribution is about $\sim 30$\,mag, which is rather faint for current telescopes, and the value of the flux in the redder band will be higher, partly caused by the cosmic reddening effect. 

The multiple images of a lensed kilonova (and the afterglow discussed later) are magnified or de-magnified by the intervening lens galaxy, 
{
and there may be significant time delays between these arrivals. 
} 
If the lensed images can be spatially resolved, we may obtain the LCs of different images separately. Therefore, in this paper, we define a lensed kilonova as detectable based on the following criteria: (1) 
{
the number of lensed images should be larger than one; 
} 
(2) the peak ($m_{\rm m}^{\rm peak}$) of the LC of the faintest image of the kilonova is at least $0.5$\,mag brighter than the limiting magnitude ($m_{\rm lim}$) of the searching telescope(s), i.e.,
\begin{equation}
m_{\rm m}^{\rm peak}+0.5<m_{\rm lim},
\end{equation}
\begin{equation}
m_{\rm m}^{\rm peak}=-2.5\log\left( {\min(\{|\mu_{\rm m}|\})} F_{\nu}^{\rm peak}\right)-48.6,
\end{equation}
where 
{
$F_{\nu}^{\rm peak}$ denotes the kilonova's flux in the observer frame after applying the K-correction {(by redshifting the spectral models and flux)} but without lensing magnification, and $\{|\mu_{\rm m}|\}$ is the absolute value list of the magnification factors of the multiple images.}  
{Notably, MLG+23 found that the peak duration of a typical lensed kilonova LC $\Delta T_{\rm p-0.5}$ with magnitude $m_{\rm m}^{\rm peak}$ to $m_{\rm m}^{\rm peak}+0.5$ is sufficiently larger than $1$ day for optical and infrared bands. Therefore, this criterion (2) allows several independent observations to characterize the light curve's rapid evolution, for telescopes have sufficient warning to schedule observations at peak brightness }
; (3) the angular separation of the {multiple} images $\Delta \theta$ is required to be larger than $\sim 0.12''$, as the designed angular resolution of the F-band in RST,
{
so that the LCs of multiple images remain distinct and do not overlap.
}

\subsection{Lensed sGRBs}
\label{sec:grb}

The {prompt emission of sGRB} is often modeled by a Gaussian-structured jet \citep[e.g., ][]{2018ApJ...867...57R}, of which the Lorentz factor $\Gamma_{\theta}$ is assumed to be
\begin{equation}
\Gamma_\theta \beta_\theta=\Gamma_0 \beta_0 \exp \left(-\theta^2 / \theta_{\mathrm{c}}^2\right),
\end{equation}
where $\beta_{\theta}$ is the corresponding velocity, $\theta_{\rm c}$ is the structure parameter denoting for the half-opening angle of the jet. Therefore, by integrating over the azimuth angle, the isotropic equivalence energy observed with viewing angle $\theta_{\rm v}$, $E_{\mathrm{iso}}\left(\theta_{\mathrm{v}}\right)$ can be calculated by \citep[e.g.,][]{2022MNRAS.511.2356M}
\begin{equation}
E_{\mathrm{iso}}\left(\theta_{\mathrm{v}}\right)=\pi E_{\mathrm{tot}, \gamma} \int_0^{\theta_{\mathrm{w}}} \mathrm{d} \theta \frac{\left(2 a^2+b^2\right) \epsilon(\theta) \sin \theta}{\Gamma^4(\theta)\left(a^2-b^2\right)^{5 / 2}},
\label{eq:iso}
\end{equation}
with 
\begin{equation}
\epsilon(\theta)=\frac{\exp \left(-\theta^2 / \theta_{\mathrm{c}}^2\right)}{\pi \theta_{\mathrm{c}}^2\left[1-\exp \left(-\theta_{\mathrm{w}}^2 / \theta_{\mathrm{c}}^2\right)\right]},
\end{equation}
where $\theta_{\rm w}$ is the truncated angle,  $a(\theta,\theta_{\rm v})=1-\beta_{\theta}\cos{\theta_{\rm v}}\cos{\theta}$ and  $b(\theta,\theta_{\rm v})=\beta_{\theta}\sin{\theta_{\rm v}}\sin{\theta}$. It can be seen from Equation~\eqref{eq:iso}, the total $\gamma$-ray energy  $E_{\mathrm{tot}, \gamma}$ is a key parameter to determine the prompt-emission, which is generated according to Appendix. 

When computing the prompt fluence $F_{\gamma}(\theta_{\rm v})$ of the sGRBs, the K-correction ($k$) and bolometric-correction ($B$) are requisite, and
\begin{equation}
F_{\gamma}(\theta_{\rm v})=\frac{E_{\rm iso}(\theta_{\rm v})}{4\pi d^2_{L}}\times  B\times k(z_{\rm s}), 
\end{equation}
which can be estimated by the Band-function \citep[e.g.,][]{1993ApJ...413..281B}. Here the bolometric correction is conducted within the range of $10-1000$\,keV with respect to the broadband $1-10^4$\,keV. Note that the correction never excesses an order of magnitude to the observed fluence. We assume that $\Gamma_0$ is uniformly distributed within the range of $[100,800]$, $\theta_{\rm c}$ follows a log-normal distribution $P(\theta_{\rm c}) \propto 1/(\sigma\theta_{\rm c})\exp{[-(\ln{\theta_{\rm c}}-\mu)^2/(2\sigma^2)]}$ \footnote{In the unit of degree, $\mu=1.742$ and $\sigma=0.916$ \citep{2013MNRAS.428.1410G}}  and $\theta_{\rm w}=2\theta_{\rm c}$ \citep[e.g.,][]{2012ApJ...746...48M}.

\begin{figure}
\centering
\includegraphics[width=1.0\columnwidth]{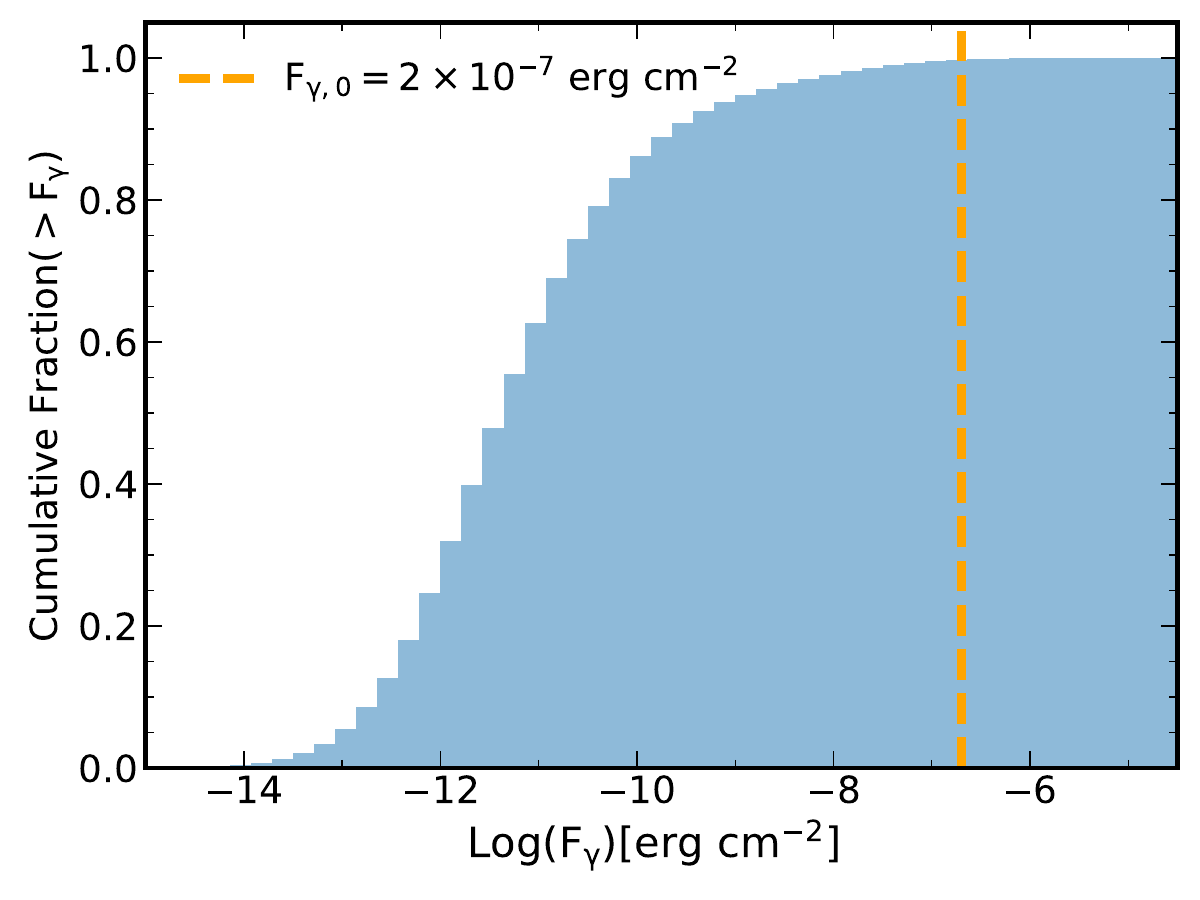}
\caption{Cumulative distribution of the prompt fluence $F_{\gamma}$ of the mock sGRBs with redshift range of $[0,5)$. The orange dashed line show the detection {threshold for Fermi-GBM, i.e., $F_{\gamma,0}=2\times 10^{-7} \rm ~erg~cm^{-2}$}.
}
\label{fig:fluence}
\end{figure}

Since the time-delay between multiple images are much longer than the duration of a single sGRB (normally $T_{90}<2 \rm s$), lensed sGRB signals {may be} identified in a time sequence. Based on this consideration, we set the criterion for detecting lensed sGRBs as $|\mu_{\rm m}|\times F_{\gamma}(\theta_{\rm v})> F_{\gamma,0}$, where $F_{\gamma,0}=2\times 10^{-7}\rm erg cm^{-2}$ is the detection limit for Fermi-GBM. Figure~\ref{fig:fluence} shows the cumulative distribution of the prompt fluence $F_{\gamma}$ of the mock sGRBs. As seen in this figure, only a small fraction, approximately $0.12\%$ of sGRBs can exceed the detection threshold for Fermi-GBM. By multiplying the total number of BNS mergers by the detection sky coverage of Fermi-GBM (half sky), 
{
we estimate the total detection rate of sGRBs from the mock BNS mergers to be around {tens to hundreds} per year,
}
which is consistent with the result obtained in \cite{2021ApJ...916...54Y}. 

\subsection{Lensed afterglows}
\label{sec:afterglow}

The LCs of afterglows for each sGRB produced by BNS mergers can be generated with the popular Python Package {AfterglowPy} \citep{2020ApJ...896..166R}, given a certain total kinetic energy $E_{\rm tot, K}$ of the relativistic jet, viewing angle and physical properties of the interacting interstellar medium (ISM). 
{
In this work, we compute $E_{\rm tot, K}$ for each mock BNS merger, parameterized by its different component masses, by assuming the relativistic jet is produced via the Blandford-Znajek mechanism (see Appendix~\ref{app:jet} for details). By sampling a range of viewing angles, we may map out the resulting diversity in afterglow emission from BNS mergers across the universe, extending our analysis well beyond the GW170817 event. 
}
As for ISM properties, including the number density of electrons $n_{{\rm e},0}$, power law index of accelerated shock $p$, the accelerated electron energy fraction $\epsilon_{\rm e}$ and the magnetic field energy fraction $\epsilon_{e,B}$,  we refit the LC of GW170817 afterglow with the full uniform dataset by \citet{Makhathini:2020ece}. Then, we assume the fitted parameters, i.e.,   $n_{e,0}$,  $p$,  $\epsilon_{e}$ and $\epsilon_{e,B}$ to be the same across all the mock afterglow samples. More details on the fitting procedure and the posteriors can be seen in Appendix~\ref{app:Fit}. { We also note that the response time will affect the detection rate of lensed afterglow signals. While rapid-response missions such as the Einstein Probe (EP; \citealt{2025SCPMA..6839501Y}) and robotic ground telescopes like GOTO \citep{2024SPIE13094E..1XD} can initiate observations within minutes under ideal conditions, the average latency in realistic large-scale campaigns is typically longer. During the first half of the LVC O3 run, GOTO-4 reported a mean first-observation time of 9.90 hours post-trigger, despite its sub-minute response capability in specific instances \citep{2020MNRAS.497..726G}. Similarly, in the recent O4a run, ZTF followed up GW candidates with potential neutron star components with latencies of several hours, such as the 10-hour delay for GW230529 \citep{2024PASP..136k4201A}. Therefore, we adopt a 10-hour response time as a representative and conservative baseline. This accounts for the logistical latency inherent in high-confidence GW parameter estimation and the scheduling constraints of next-generation deep-field facilities such as RST and ATHENA. For events captured earlier by fast-responding facilities, our estimated rates effectively serve as a robust lower bound, for example, adopting a shorter response time of 1 hour would enhance the detection rates by a factor of $\sim 2-3$. }

\begin{figure*}
\includegraphics[width=0.65\columnwidth]{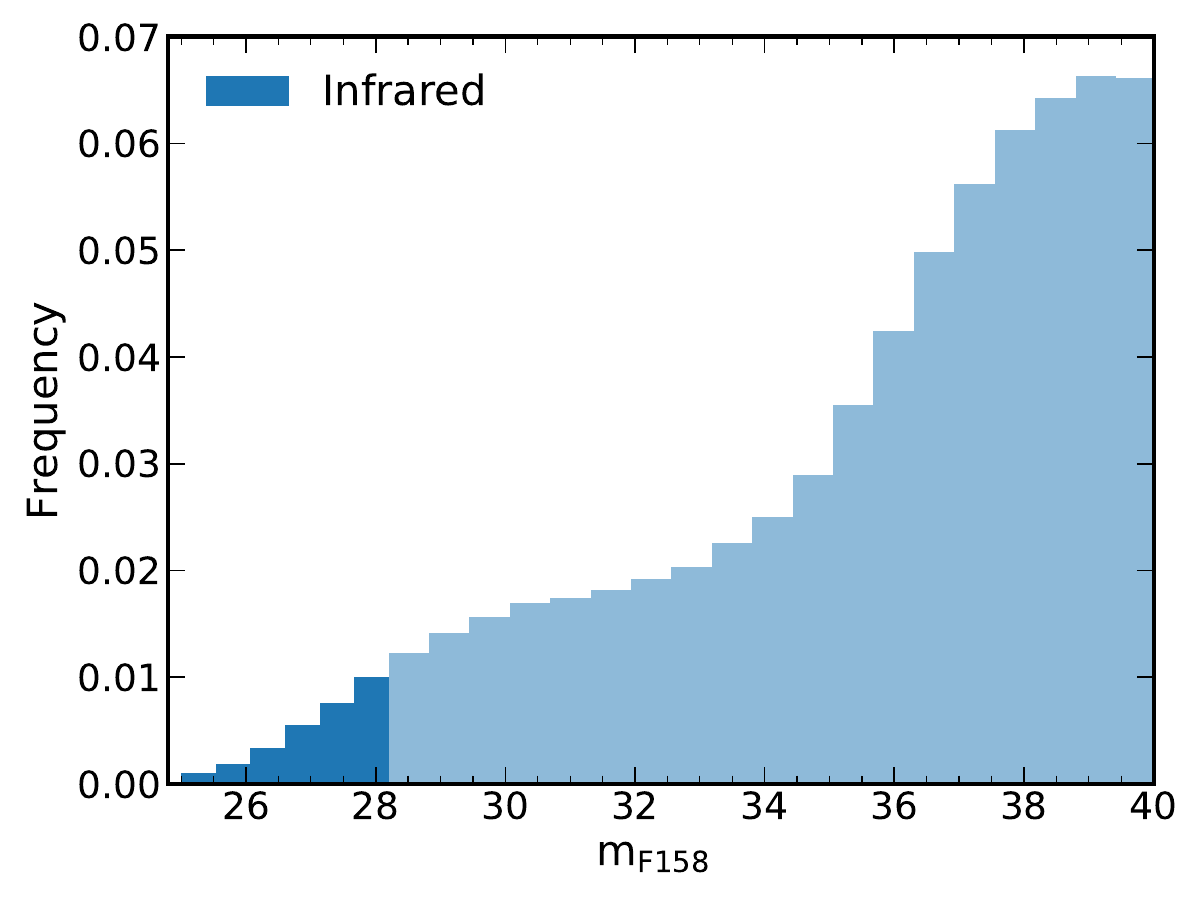}
\includegraphics[width=0.65\columnwidth]{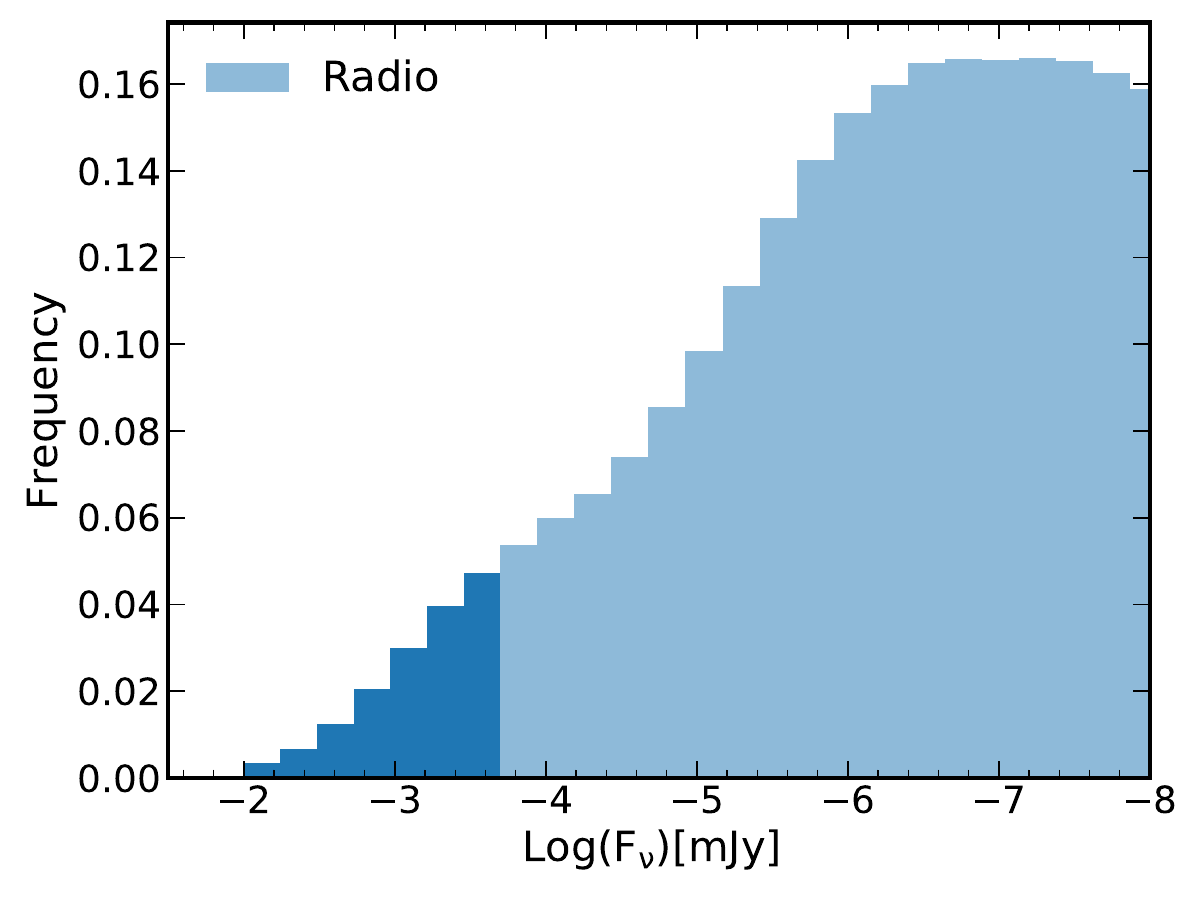}
\includegraphics[width=0.65\columnwidth]{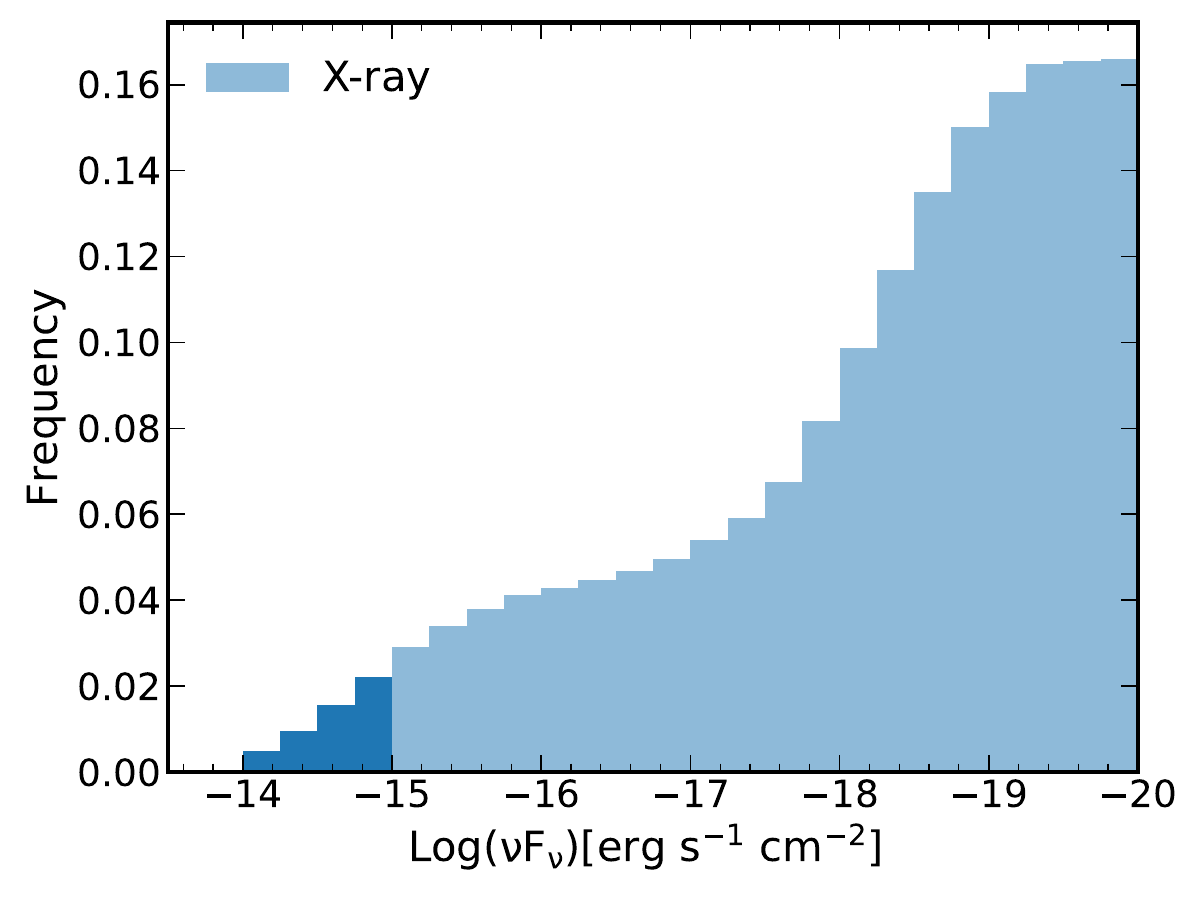}
\centering
\caption{
The normalized distribution of the peak flux of afterglow signals after $10$\,h {within the redshift range of $[0,5)$}, in the infrared F158 (left), radio $8.4$\,GHz (mid), and X-ray $5$\,keV (right) bands, respectively. {The deeper color region in each panel shows the population below the detection threshold of RST, Chandra and SKA respectively. }
}
\label{fig:aft}
\end{figure*}

Figure~\ref{fig:aft} shows the peak flux distribution of the mock afterglow signals in different bands. The infrared band is set to be the F158 band of RST, while the radio and X-ray bands are set to be $8.4$\,GHz and $5.0$\,keV, respectively, which are the designed working frequencies of the Squared Kilometer Array (SKA) and Chandra X-ray telescope \citep[e.g.,][]{2019MNRAS.489.1919T}. In our sample, the GW170817-like afterglow is rare. This is mainly due to that most BNS mergers are much more distant than GW170817. In addition, the distribution of $E_{\rm tot, K}$ of our mock BNS merger samples, has a median value of $\sim 1.43\times10^{49}$\,erg, substantially smaller than the total kinetic energy $E_{\rm tot, K}$ of GW170817 (approximately $\sim 3.5\times10^{49}$,\,erg). {Notably, we also find that applying different cuts (e.g.,$\sim 10$ days) does not significantly alter the peak flux distribution shown in Figure~\ref{fig:aft}. This is primarily because the majority of the lensed BNS afterglows in our sample are off-axis due to the collimated nature of the jets, as the case of GW170817 \citep[e.g.,][]{Makhathini:2020ece}. The LC of off-axis afterglows typically reach their flux peaks several days to months after the merger and then exhibit a power law decay, leading to a similar distribution though with different specific choice of the early-time cut.}

For a consistency check, we also estimate the LVK O4 joint detection rate of the GW signals from BNS mergers and the associated afterglow signals in each band. Specifically, we find $\dot{N}_{\rm LVK+AG}\sim 0.51/0.25/0.52$\,yr$^{-1}$ for the infrared, radio and X-ray bands, assuming the detection thresholds of $m_{\rm lim}=28$\,mag, $F_{\nu}=0.1$\,mJy, and $\nu F_{\nu}=10^{-15}$\, ergs$^{-1}$\,cm$^{-2}$, respectively. These values are almost the same as those given in \citet[][see Table 1 therein]{2022ApJ...937...79C} under their adopted normalization. This consistency validates the use of the mock samples generated in this paper to estimate the lensed afterglow signals.

\subsection{Lensed host galaxies}
\label{sec:host}
To estimate the detection probability of the lensing signatures for the host galaxies of lensed BNS events $P({\rm H}|{\rm GW})$, we adopt the following Bayesian formalism in our previous work \citep[e.g., see][]{2022ApJ...940...17C}
\begin{equation}
\label{bayes}
P({\rm H}|{\rm GW})=\frac{P({\rm GW}|{\rm H})P({\rm H})}{P({\rm GW}|{\rm H})P({\rm H})+P({\rm GW}|{\rm H^{\ast}})P({\rm H^{\ast}})},
\end{equation}
where ``H'' and ``GW'' denote host galaxies with identifiable lensing images and detected GW events identified as lensed, respectively, ${\rm H}^\ast$ represents lensed galaxies without identifiable lensing images, and the probabilities $P({\rm GW}|{\rm H})$, $P({\rm H})$, $P({\rm GW}|{\rm H}^{\ast})$, and $P({\rm H}^{\ast})$ represent the probability of lensed host galaxies with identifiable lensing images that have identified lensed GW events, the probability of lensed galaxies that have detectable lensing signatures, the probability for lensed galaxies without detectable lensing signatures that have detected lensed GW events, and the probability of lensed galaxies without detectable lensing signatures, respectively. 
{
To calculate the above terms, we begin by generating mock host galaxies from the sky catalog of \citet{2010SPIE.7738E..1OC}, assigning each galaxy intrinsic properties, apparent magnitude $m_r$, $m_i$, $m_z$ in the r, i and z bands, stellar mass $M_{\ast}$, ellipticity $q_{\rm s}$, and effective radius. Then, we populate randomly these hosts with GW-emitting BNS mergers, drawing the number of mergers per galaxy in proportion to their stellar mass $M_{\ast}$. We assume within each host, mergers trace the galaxy's surface brightness profile, reflecting their origin in massive-binary evolution. Furthermore, we assume that the future galaxy-galaxy lensing catalog will be assembled by space-borne telescopes, such as Euclid \citep[e.g.,][]{2013LRR....16....6A}, the Chinese Space Station Telescope (CSST) \citep[e.g.,][]{2019ApJ...883..203G}, and RST. For more details of our sampling procedure, see \citet{2022ApJ...940...17C}.
}

In this paper, we adopt the following criteria for identifiable lensed host galaxies designed by \citet{Collett_2015} \citep[see also][]{2022ApJ...940...17C,2024MNRAS.533.1960C}

\begin{enumerate}
\item [1)] The center of the host galaxy should be within the 
{
Einstein light cone of foreground galaxies}, thus {may produce multiple images}, i.e., $x_{\rm s}^2+y_{\rm s}^2\lesssim \theta_{\rm E}^2$, where $\theta_{\rm E}$ is the Einstein radius.
\item [2)] The image and counter-image must be resolvable, i.e., $R_{\rm e}^2+s^2/2\lesssim \theta_{\rm E}^2$, where $s$ is the seeing. For example, $s$ is $0.18^{\prime\prime}$ for the i-band of CSST \citep[e.g.,][]{2019ApJ...883..203G}, $0.23^{\prime\prime}$ for the VIS-band of Euclid \citep[e.g.,][]{2013LRR....16....6A}, and $0.11^{\prime\prime}$ for the J-band of RST \citep[e.g.,][]{2013arXiv1305.5422S}.
\item [3)] {The tangential stretching of the source should be detectable}, i.e., ${\mu_{\rm tot}R_{\rm e}>s}$, where $\mu_{\rm tot}$ is the total magnification of the source. 
\item [4)] {The total S/N of the lensed images (determined by calculating the quadrature sum of all pixels in the lensed images with an SNR greater than 1) should be large enough for the identification of lensed images, i.e., ${\rm S/N} > 20$.}
\end{enumerate}
For the EM observations searching for the lensed hosts, {we approximate the VIS-band magnitude of Euclid by the mean apparent magnitude of host galaxies in the $r$, $i$ and $z$ bands, i.e., $ m_{\rm VIS} = (m_{\rm r}+m_{\rm i}+m_{\rm z})/3$  \footnote{{The accurate expression should be logrithmic mean rather than arithmetic mean. However, using the observed photometry of the host galaxy of GW170817, i.e., NGC 4993 ($r=12.16, i=11.81, z=11.57$; \citealt{2017ApJ...848L..22B}) as a representative template, this approximation introduces a bias of only $\sim 0.03$ mag compared to a rigorous flux-weighted average, which will only affect our estimation slightly.}} 
\citep[see the same approximation in][]{Collett_2015}, and the J-band magnitude of RST by the extrapolation made from SDSS values for the $i$ and $z$ bands, i.e., $ m_{\rm J}=m_{\rm z}-4.4(m_{\rm i}-m_{\rm z})$\footnote{{While our $J$-band scaling relation is derived from less sensitive SDSS data, it provides a robust first-order approximation predicated on the universality of stellar population scaling laws.  A validation test using NGC 4993 yields a predicted magnitude  within $\sim 0.4$ mag of the observed value, which may lead to a $\sim 10\%$ variation in the estimation on $P\rm (H\mid GW)$.}}\citep[see the approximation in][]{2020RNAAS...4..190W} for the mock host galaxies.  } 

\section{Joint detection rate of lensed GW + sGRB}
\label{sec:gw+sgrb}

\begin{figure}
\includegraphics[width=1.0\columnwidth]{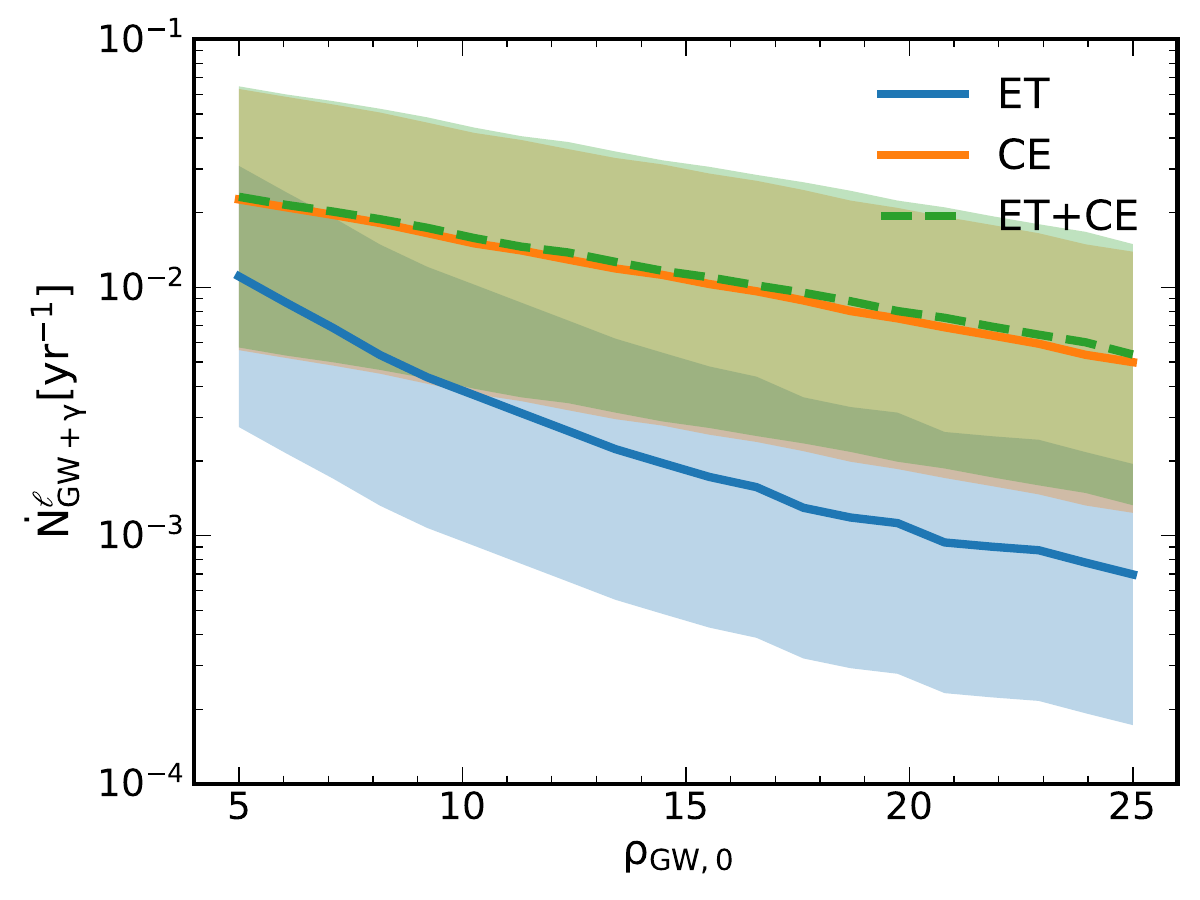}
\includegraphics[width=1.0\columnwidth]{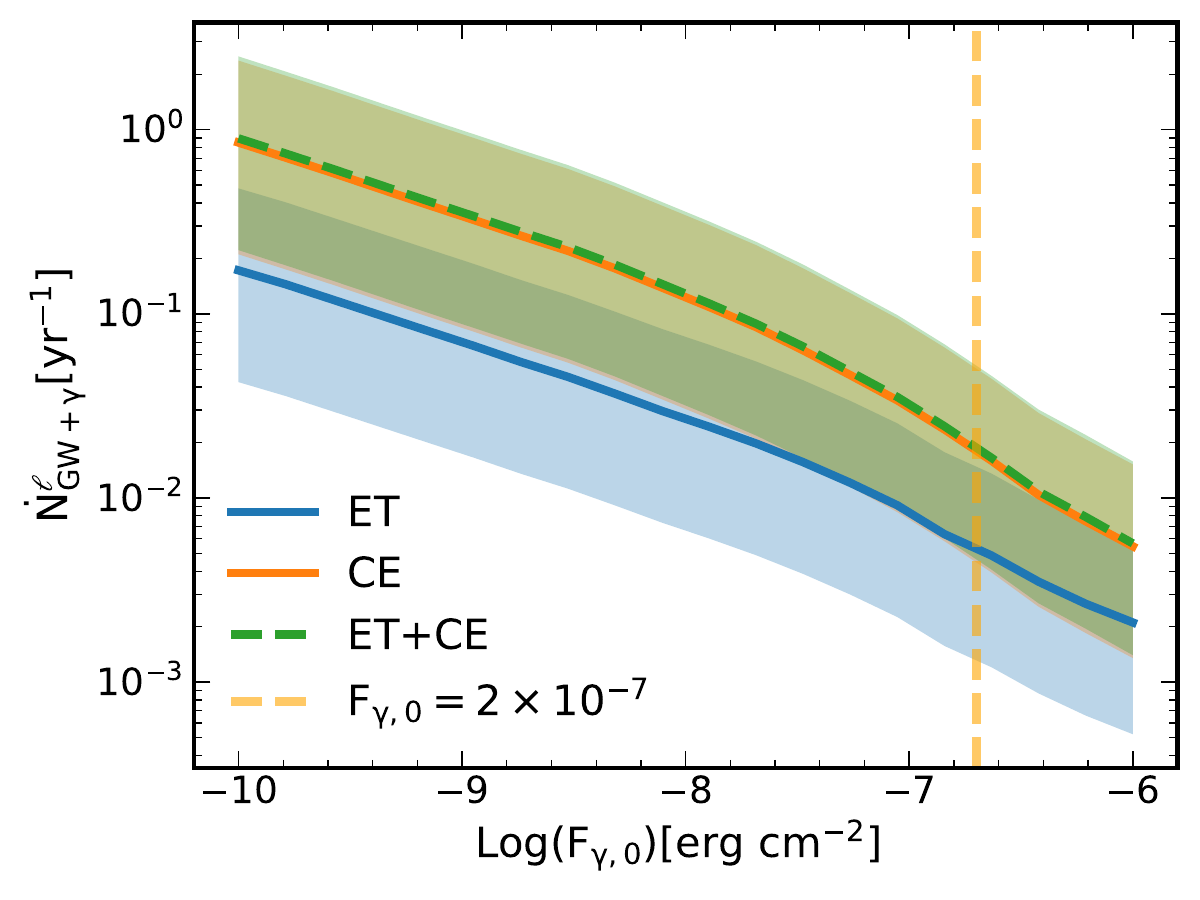}
\centering
\caption{
The {total} joint detection rate of lensed GW and sGRB signals per year by third generation GW detectors and Fermi-GBM. The blue, orange and green solid lines show the results of ET, CE and their network respectively. In the top panel, the detection threshold for sGRBs is fixed as $F_{\gamma,0}=2\times 10^{-7}$\,erg~cm$^{-2}$, which is the detection threshold of Fermi-GBM. In the bottom panel, the detection S/N threshold of GW signals is set to be $\varrho_{\rm GW,0}=8$ and the yellow dashed line show the $F_{\gamma,0}=2\times 10^{-7}$\,erg~cm$^{-2}$ case. 
}
\label{fig:sgrb+gw}
\end{figure}

{Because GW and sGRB detectors operate as blind-trigger facilities, lensed multi-messenger events must be identified via posterior cross-matching of independent triggers. Thus, the search strategy is predicated on the lensing time delay $\Delta t$, where the late-arriving image acts as a temporal trigger to search the preceding GW database.} Considering this strategy, the total detection rate $\dot{N}^{\ell}_{\rm GW+sGRB}$ is determined by the S/N threshold of the GW detection, $\rho_{\rm GW,0}$, and the fluence threshold of the $\gamma$-ray telescope, $F_{\gamma,0}$.
We present our main results on this dependence in Figure~\ref{fig:sgrb+gw}. The top panel of Figure~\ref{fig:sgrb+gw} shows the dependence of $\dot{N}^{\ell}_{\rm GW+sGRB}$ on $\varrho_{\rm GW,0}$ with different third-generation GW detectors and the network composed by them, assuming $F_{\gamma,0}=2\times 10^{-7}$\,erg~cm$^{-2}$. On one hand, the higher the $\varrho_{\rm GW,0}$, the lower the $\dot{N}^{\ell}_{\rm GW+sGRB}$ due to the limited sensitivity of GW detectors. On the other hand, given the same $\varrho_{\rm GW,0}$, the joint detection rate of CE will be higher than that of ET due to its significantly greater sensitivity. For example, if we adopt the typical threshold $\varrho_{\rm GW,0}=8$, {the total joint detection rate are about $\sim0.006$, $0.018$, and $0.019$
per year} for ET, CE, and CE+ET, 
{
corresponding to a rough relative fraction of $\sim 2.6\times 10^{-3}$ among all detectable lensed BNS merger GW signals. This may be attributed by the following reasons. On the one hand, only a fraction of BNS mergers may produce jet with substantial energy to break out from the mass ejecta and thus produce sGRB signals. On the other hand, the sGRB signals are highly collimated compared with GW signals and thus subject to strong selection effect.}

Due to the rarity of such detectable events, in the following analysis on lensed multi-messenger signals from BNS mergers, we do not discuss the detection rate of lensed sGRBs. Nevertheless, we note that with the development of future $\gamma$-ray telescopes, $F_{\gamma,0}$ may be significantly lower than $2\times 10^{-7}$\,erg~cm$^{-2}$, potentially leading to a substantially higher joint detection rate. The bottom panel of Figure~\ref{fig:sgrb+gw} shows the dependence of $\dot{N}^{\ell}_{\rm GW+sGRB}$ on $F_{\gamma,0}$, assuming $\varrho_{\rm GW,0}=8$. As seen in the figure, the expected joint detection rate $\dot{N}^{\ell}_{\rm GW+sGRB}$ can exceed $0.1$\,yr$^{-1}$ assuming  $F_{\gamma,0}$ is sufficiently lower than $\sim 1.5\times 10^{-8}$\,erg~cm$^{-2}$ for CE or the ET+CE network . Only when the sensitivity of $\gamma$-ray detectors is {higher than that of Fermi-GBM by a factor of $\gtrsim 12$}  can one detect a lensed sGRB event within a ten-year observational period. 
{
In addition, double-image cases dominate the total detection rate, while triple/quadruple cases contribute only marginally. This pattern holds for both lensed kilonovae and afterglows, which is evident in Table~\ref{table:rate}.
}

{
It is important to emphasize that the exceedingly low joint detection rate $\dot{N}^{\ell}_{\rm GW+sGRB}$ we derived does not include the practical hurdles of pairing lensed sGRB signals and GW triggers in practical data analysis \citep[e.g.,][]{2022ApJ...924...49C,2025RSPTA.38340122L}. In particular, the large localization uncertainties for both sGRBs and GWs \citep[e.g.,][]{2015ApJS..216...32C} elevate the false-alarm probability and make confident associations far more difficult. Consequently, the joint detection of strongly lensed GW-sGRB events will remain an observationally inefficient endeavor. 
}

\begin{table*}
\caption{{The joint detection rate of lensed GW and multi-messenger signals of BNS mergers with ET-CE network, assuming the GW detection threshold to be $8$. The first column shows the band adopted to detect lensed multi-messenger signals. The second, third and fourth columns show the joint detection rate of double, triple and quadruple image cases, respectively. 
{The fifth and sixth columns show the detection rate of unlensed multi-messenger signals and the relative fraction of lensed multi-messenger signals among all lensed detectable BNS GW events.} All uncertainties are the direct consequence of the uncertainty {in the local merger rate density from GWTC-4, i.e., $89_{-67}^{+159} \rm Gpc^{-3} yr^{-1}$.} The rates presented here may be subject to some additional uncertainties due to merger rate density evolution, the detector downtime and also the radiation models. Thus, we also report the relative fraction of detectable lensed events among all detectable events, which we expect to be subject to less uncertainty \citep[e.g.,][]{2021ApJ...921..154W}. } }
\label{table:rate} 
\centering
\begin{tabular}{lcccccc} \hline    \hline
        & Band & Double & Triple & Quadruple & Unlensed events & Relative fraction  \\
       \hline
        sGRB  & $\rm 10-1000keV^{(1)}$  & $0.015^{+0.027}_{-0.012} \rm yr^{-1}$ & $0.0020_{-0.0015}^{+0.0035} \rm yr^{-1}$ & $0.0017_{-0.0013}^{+0.0030} \rm yr^{-1}$ & $6.8^{+12.1}_{-5.1}\times10^{1} \rm yr^{-1}$ & $2.6\times 10^{-3}$  \\ \hline
        Kilonova & $\rm F106^{(2)}$ & $0.35^{+0.63}_{-0.27} \rm yr^{-1}$ & $0.050^{+0.089}_{-0.038} \rm yr^{-1}$ & $0.047^{+0.084}_{-0.036}\rm yr^{-1}$ & $5.1^{+9.1}_{-3.9}\times10^{3}\rm yr^{-1}$ & $3.2\times 10^{-2}$ \\ 
         & F158 &$0.42_{-0.32}^{+0.75}\rm yr^{-1}$&$0.070^{+0.124}_{-0.052}\rm yr^{-1}$&$0.058^{+0.104}_{-0.044} \rm yr^{-1}$&$5.5^{+9.8}_{-4.1}\times10^{3}\rm yr^{-1}$ & $3.9\times 10^{-2}$\\
         & F213 &$0.058^{+0.104}_{-0.044}\rm yr^{-1}$ &$0.013^{+0.023}_{-0.0096}\rm yr^{-1}$ & $0.0058^{+0.010}_{-0.0044}\rm yr^{-1}$ &$3.6^{+6.4}_{-2.7}\times10^{2} \rm yr^{-1}$  & $5.5\times10^{-3}$ \\ \hline
        Afterglow & F158 & $0.012^{+0.022}_{-0.009}\rm yr^{-1}$ &$0.0018^{+0.0033}_{-0.0014}\rm yr^{-1}$& $0.0018^{+0.0032}_{-0.0014}\rm yr^{-1}$ & $3.8^{+6.8}_{-2.8}\times10^{2}\rm yr^{-1}$ & $1.1\times10^{-3}$\\
         & $\rm 5keV^{(3)}$ & $0.019^{+0.035}_{-0.015}\rm yr^{-1}$ & $0.0024^{+0.0043}_{-0.0018}\rm yr^{-1}$ &$0.0027^{+0.0048}_{-0.0020} \rm yr^{-1}$ & $5.6^{+10.0}_{-4.2}\times10^{2}\rm yr^{-1}$ & $1.8\times10^{-3}$\\
         & $8.4\rm GHz^{(4)}$ & $0.056^{+0.099}_{-0.042}\rm yr^{-1}$&$0.0050^{+0.0089}_{-0.0038}\rm yr^{-1}$ &$0.0070^{+0.012}_{-0.0052}\rm yr^{-1}$ & $1.6^{+2.8}_{-1.2}\times10^{3}\rm yr^{-1}$ & $4.9\times 10^{-3}$ \\
        \hline\hline
\end{tabular}
\tablecomments{(1) The detection threshold for Fermi-GBM is assumed to be $2\times 10^{-7}~\rm erg~cm^{-2}$; (2) The limiting magnitude of RST-like filters F106/F158/F213 are assumed to be $28.2/28.1/26.2$ mag respectively; (3) The detection threshold for Chandra is assumed to be $10^{-15}\rm ~ergs^{-1}cm^{-2}$; (4) The detection threshold for SKA2 is assumed to be $0.2\nu\rm Jy$.}
\end{table*}

\section{Joint detection rate of lensed GW + Kilonovae/Afterglow}
\label{sec:gw+multi}

\subsection{ Searching strategy}

{The identification of {kilonovae and afterglows associated with} GW triggers has relied on both wide-field Target of Opportunity (ToO) observations and targeted observations of known galaxies within the GW localization region \citep[e.g.,][]{2017ApJ...848L..12A}. In the case of GW170817, wide-field searches were crucial for the follow-up campaign, while pointed galaxy-targeted observations, including those by the 1M2H team, also played an important role in the discovery and early follow-up of the optical counterpart \citep[e.g.,][]{2017Sci...358.1556C,2017Sci...358.1570D,2017ApJ...848L..26S}.}
In the coming years, the Vera C. Rubin Observatory (Rubin) will be transformational in this domain, utilizing its large field of view (FOV) and deep limiting magnitudes to perform rapid tiling of GW localization areas to search for lensed transients \citep[e.g.,][]{2024arXiv241104793A,2025RSPTA.38340118R}.
While such wide-field surveys are highly effective, the faintness of lensed BNS signals at high redshifts presents unique challenges.
{Motivated by the success of targeted galaxy follow-up in GW170817, we consider a complementary pointed strategy for lensed BNS events.} In this strategy, pre-identified galaxy-scale lens candidates within the GW localization region are prioritized for deep, high-resolution follow-up observations, rather than tiling the full localization area.
By cross-matching GW triggers with pre-defined lensing galaxy catalogs, this approach narrows the search to a manageable number of targets, typically 1--10 galaxies per square degree. 
Although this strategy is subject to the completeness of existing lensing catalogs, {it may improve the per-lens sensitivity to faint lensed kilonovae and afterglows from distant BNS mergers.}

\begin{figure}
\centering
\includegraphics[width=0.97\columnwidth]{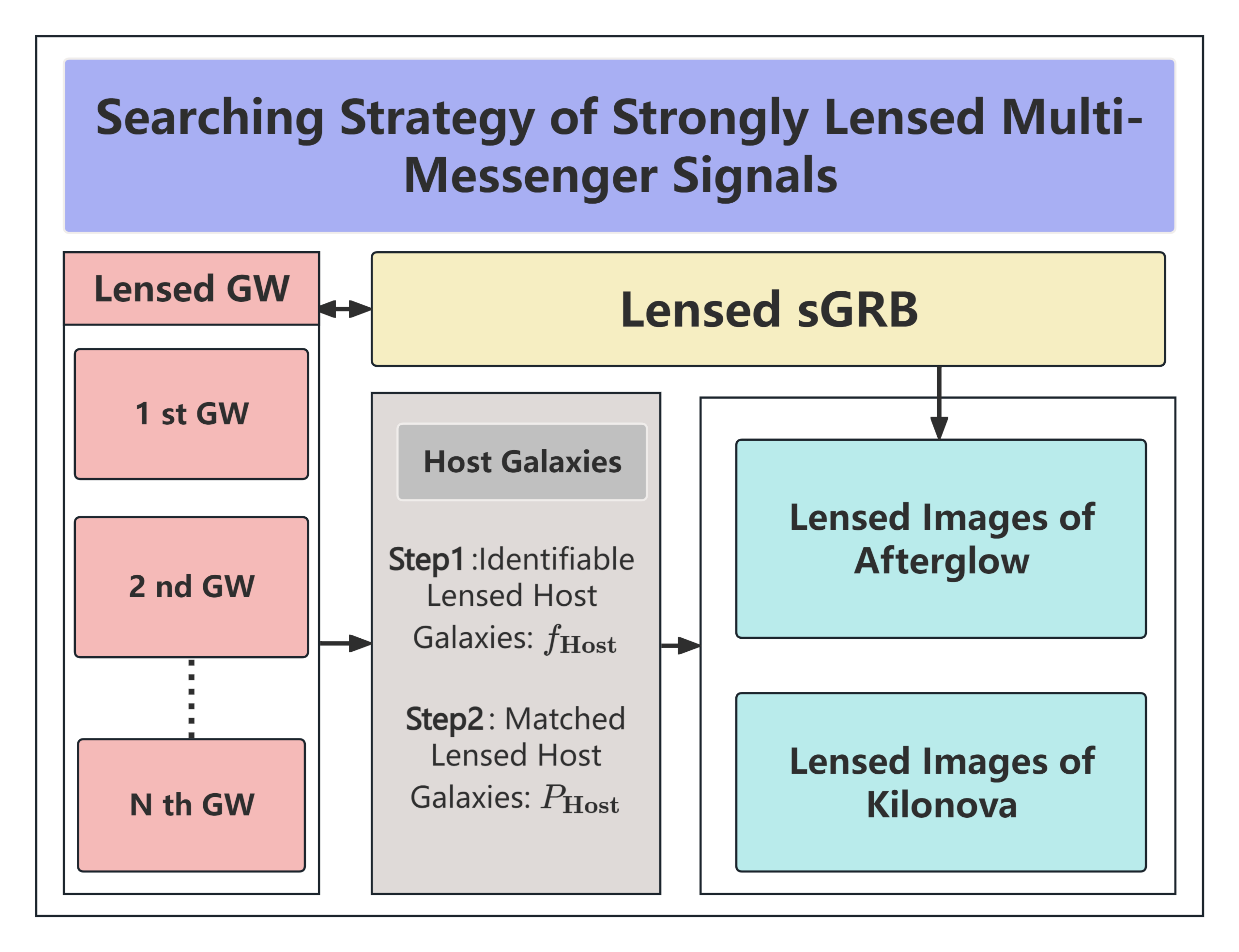}
\caption{
Steps for searching lensed multi-messenger signals of BNS mergers.  
}
\label{fig:step}
\end{figure}

In this paper, the joint detection rate is estimated using the search strategy shown in Figure~\ref{fig:step}. Once two or more GW signals are identified as multiple images of a lensed BNS merger, we use telescopes with sufficiently deep limiting magnitude to observe its lensed host galaxy, provided the host is within the galaxy-galaxy lensing catalog obtained from prior sky surveys.  

Only a fraction of host galaxies can be identified as lensed systems in a given survey, depending on the survey sky coverage and the identification criteria described in Section~\ref{sec:host}. We denote this fraction as $f_{\rm Host}$, which can be estimated as (see MLG+23)
\begin{equation}
\label{eq:host}
f_{\rm Host}=\frac{\Delta \Omega}{4\pi}\times f_{\rm OL,T}\times P({\rm H}|{\rm GW}),
\end{equation}
where $\Delta \Omega$ is the survey sky coverage, $f_{\rm OL,T}$ is the fraction of the survey area observable by the follow-up telescope at the trigger time, and $P({\rm H}|{\rm GW})$ is the conditional probability that a GW-detected lensed BNS event has identifiable host-galaxy lensing signatures. For space-borne facilities such as RST-like or Euclid-like telescopes, $f_{\rm OL,T}$ accounts for the Field-of-Regard (FOR) constraint, since only part of the survey area is observable at a given time.

After the GW event is localized, several candidate galaxy-scale lenses may lie within the sky localization region. As discussed in \citet{2004PhRvD..69b2002S} and \citet{2020MNRAS.498.3395H}, the Earth's rotation allows different lensed GW images to be observed with different detector antenna patterns. For next-generation GW detector networks, this effect can reduce the localization uncertainty of lensed GW events to $\lesssim 1\,{\rm deg}^{2}$, so that the number of candidate lens galaxies in the localization region may be only $N_{\rm cand}\sim 1$--$10$ \citep[e.g.,][]{chen25}. We therefore include a matching probability, $P_{\rm match}$, for associating the GW event with the correct lensed host. The rate of lensed GW events with detectable kilonova or afterglow emission is then
\begin{equation}
\dot{N}^{\ell}_{\rm GW+KN/AG}
=
\dot{N}^{\ell}_{\rm GW}
\times f_{\rm KN/AG}
\times f_{\rm Host}
\times P_{\rm match},
\label{eq:flen}
\end{equation}
where {$f_{\rm KN/AG}$ is the fraction of lensed kilonovae or afterglows detectable by the adopted telescope.}

{Note that the practical feasibility of the pointed follow-up can then be roughly estimated as
$T_{\rm obs}\sim N_{\rm cand}N_{\rm epoch}N_{\rm filter}t_{\rm visit}$,
where $N_{\rm epoch}$ is the number of observing epochs, $N_{\rm filter}$ is the number of filters, and $t_{\rm visit}$ is the observing time per lens per epoch/filter combination. With $N_{\rm cand}\sim 1$--$10$ and a typical value of $t_{\rm visit}\sim 1$ hour  for RST-like instruments \citep{2015arXiv150303757S}, a two-epoch single-filter confirmation would require $\sim 2$--$20$ hours. This is within or comparable to a 24-hour ToO window for favorable well-localized events, while more complete multi-epoch and multi-filter characterization would be feasible mainly for small candidate samples or shorter visits.}

\subsection{Identifiable lensed host fraction $f_{\rm Host}$}

\begin{figure}
\includegraphics[width=1.0\columnwidth]{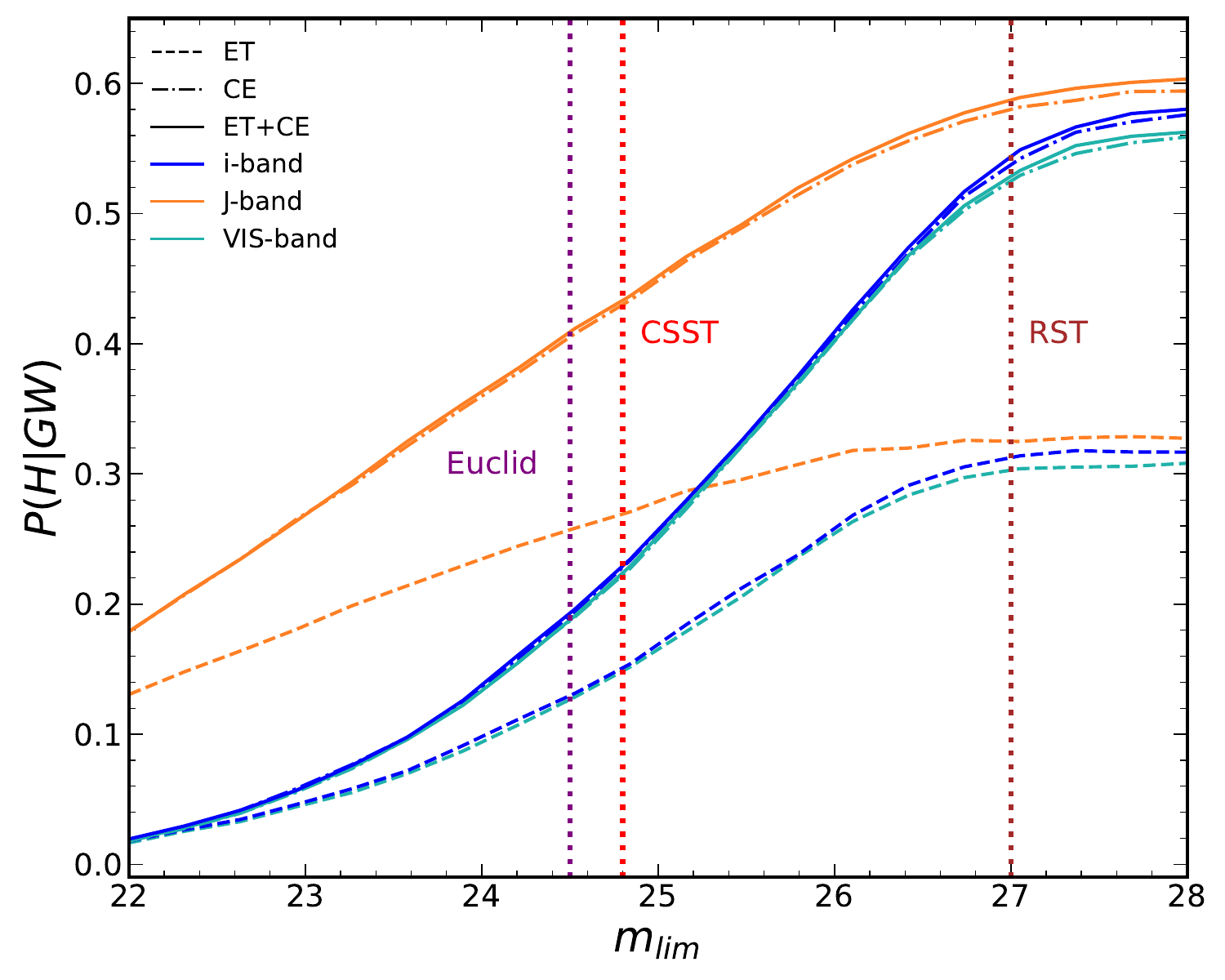}
\centering
\caption{
The probability of a lensed BNS GW event that has lensing signatures of its host galaxy (the conditional probability $P({\rm H}|{\rm GW})$) identifiable by a (survey) telescope with the limiting magnitude of $m_{\rm lim}$ for different third generation GW detectors and their network.
}
\label{fig:limit}
\end{figure}

The identifiable lensed host fraction, $f_{\rm Host}$, is composed of three terms (see Eq.~\ref{eq:host}). The first two terms, $\Delta \Omega/4\pi$ and $f_{\rm OL,T}$, depend solely on the survey itself and are independent of the properties of BNS mergers and their host galaxies. Conversely, the probability of a lensed BNS GW event exhibiting lensing signatures of its host galaxy (the conditional probability $P({\rm H}|{\rm GW})$) identifiable by a survey is related to the physical properties of the BNS merger and its host galaxy. Figure~\ref{fig:limit} presents our main results, showing the relationship between $P({\rm H}|{\rm GW})$) and the limiting magnitude $m_{\rm lim}$ for different third-generation GW detectors and their networks. It can be seen that $P({\rm H}|{\rm GW})$ increases with $m_{\rm lim}$ when $m_{\rm lim} \lesssim 26.5-27$, as the lensing signatures of lensed BNS merger hosts are more likely to be identified with increasing survey depth. However, a fraction of lensed hosts violate criteria 2) and 3) and cannot be identified, resulting in a flat $P({\rm H}|{\rm GW})$ at $m_{\rm lim} \gtrsim 27$ in this figure.

{The conditional probability $P({\rm H}|{\rm GW})$ differs between ET and CE, unlike the case for lensed stellar binary black hole mergers}. 
For example, adopting an RST-like J-band depth of $m_{\rm lim}=27$ mag for host identification, $P({\rm H}|{\rm GW})$ is approximately 30\% for ET and nearly 60\% for CE.
{This difference arises because CE can detect several times more lensed BNSs than ET \citep[e.g.,][]{2023MNRAS.518.6183M}, extending the GW-detected sample to higher redshifts where host-galaxy lensing signatures are more readily identified in redder bands.}
Moreover, $P({\rm H}|{\rm GW})$ depends on the band chosen to identify the lensing signatures of the hosts.  As shown in Figure~\ref{fig:limit}, choosing a redder band (e.g., J band instead of i band) results in a relatively higher $P({\rm H}|{\rm GW})$ at {$m_{\rm lim}$}, because lensed hosts are typically located at high redshifts and are more easily identified in redder bands than in bluer bands.

In conclusion, the identifiable lensed host fraction, $f_{\rm Host}$, is directly dependent on the filter, limiting magnitude, and survey strategy of a telescope. 
{Hereafter, we adopt a fiducial case in which the host-identification probability reaches the plateau values in Figure~\ref{fig:limit}, with $P({\rm H}|{\rm GW})\simeq 0.3$ and $0.6$ for ET and CE, respectively. Applying a representative sky-availability factor of $\Delta \Omega/{4\pi}\times f_{\rm OL,T}
\sim 0.5$ gives $f_{\rm Host}\simeq 0.15$ and $0.3$. These values correspond to a sufficiently deep lens catalog, and the rates for shallower surveys can be rescaled linearly with $f_{\rm Host}$.}


\subsection{{Lensed kilonovae/afterglow fraction: $f_{\rm KN/AG}$}}

\begin{figure*}
\includegraphics[width=0.65\columnwidth]{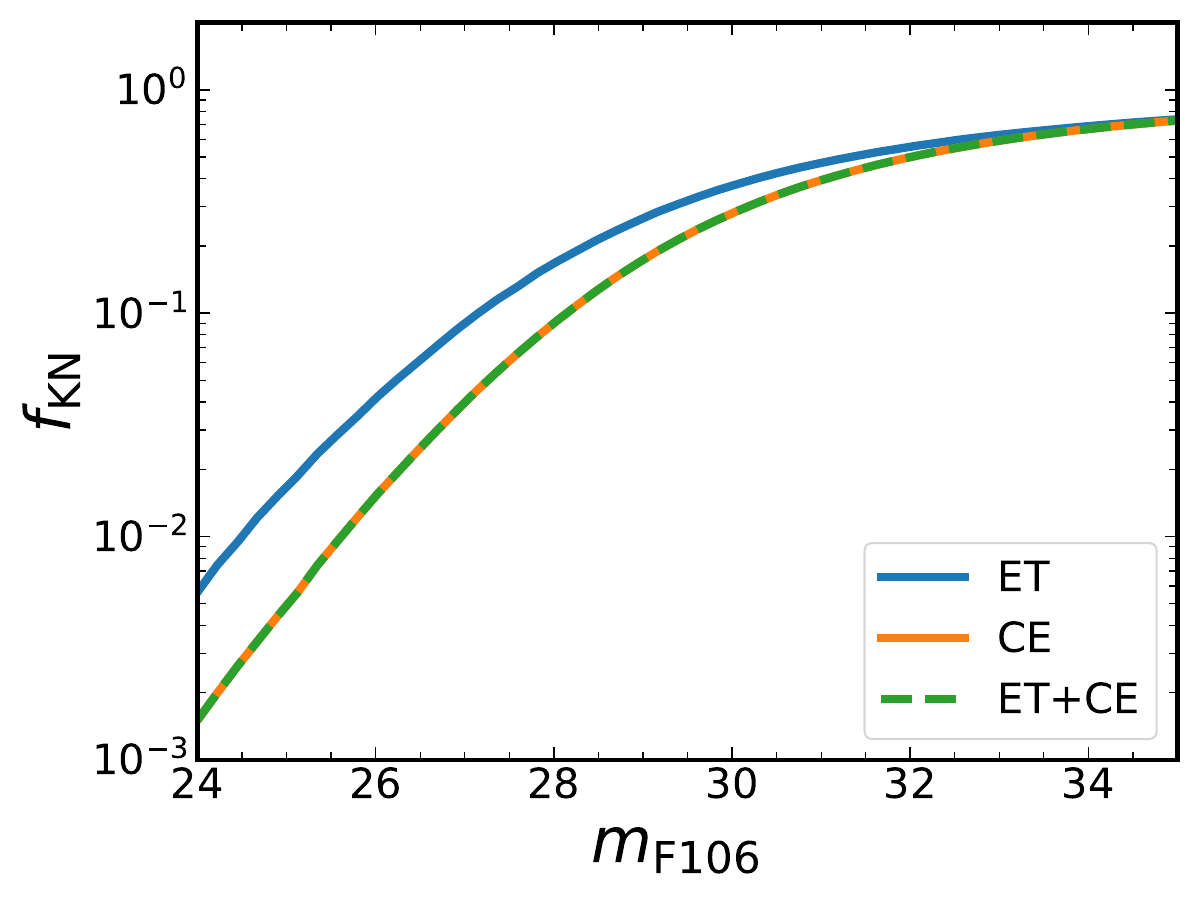}
\includegraphics[width=0.65\columnwidth]{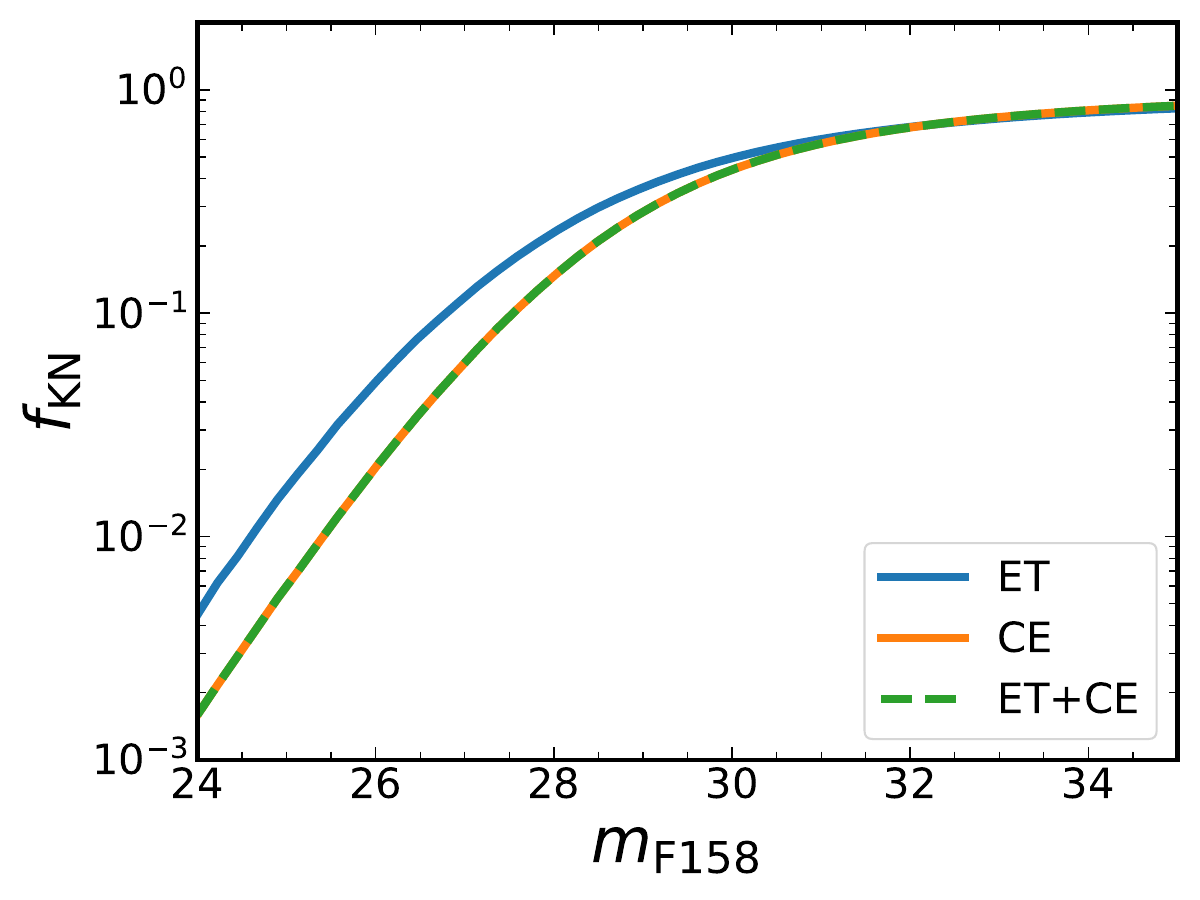}
\includegraphics[width=0.65\columnwidth]{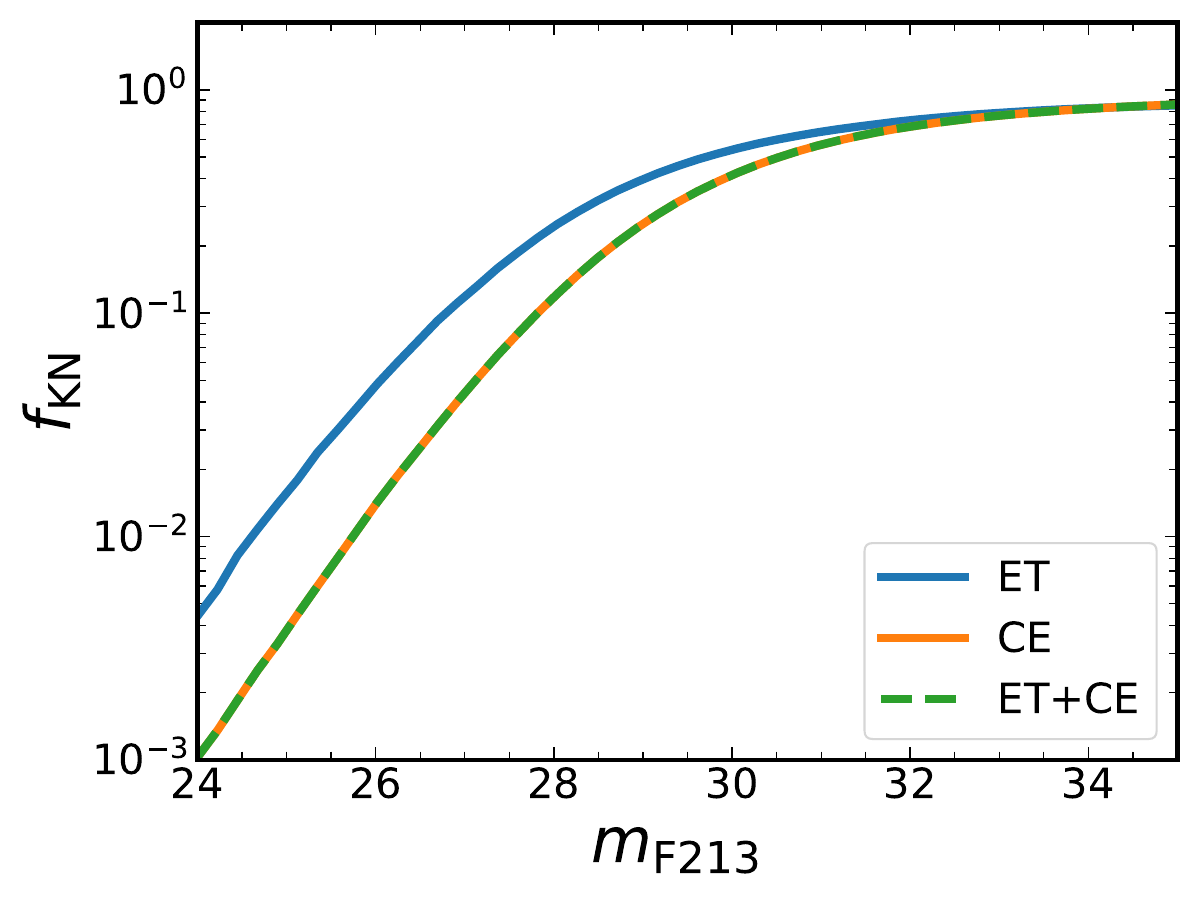}
\includegraphics[width=0.65\columnwidth]{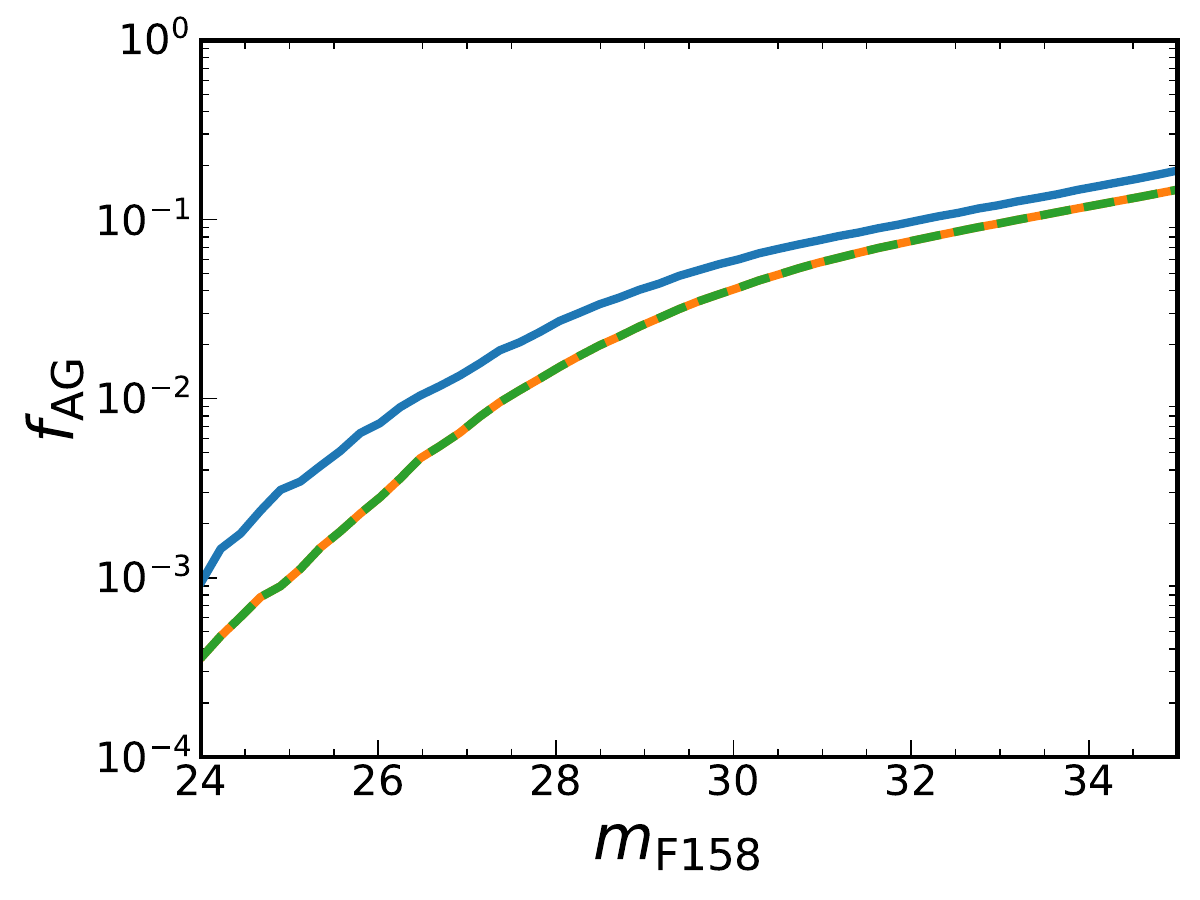}
\includegraphics[width=0.65\columnwidth]{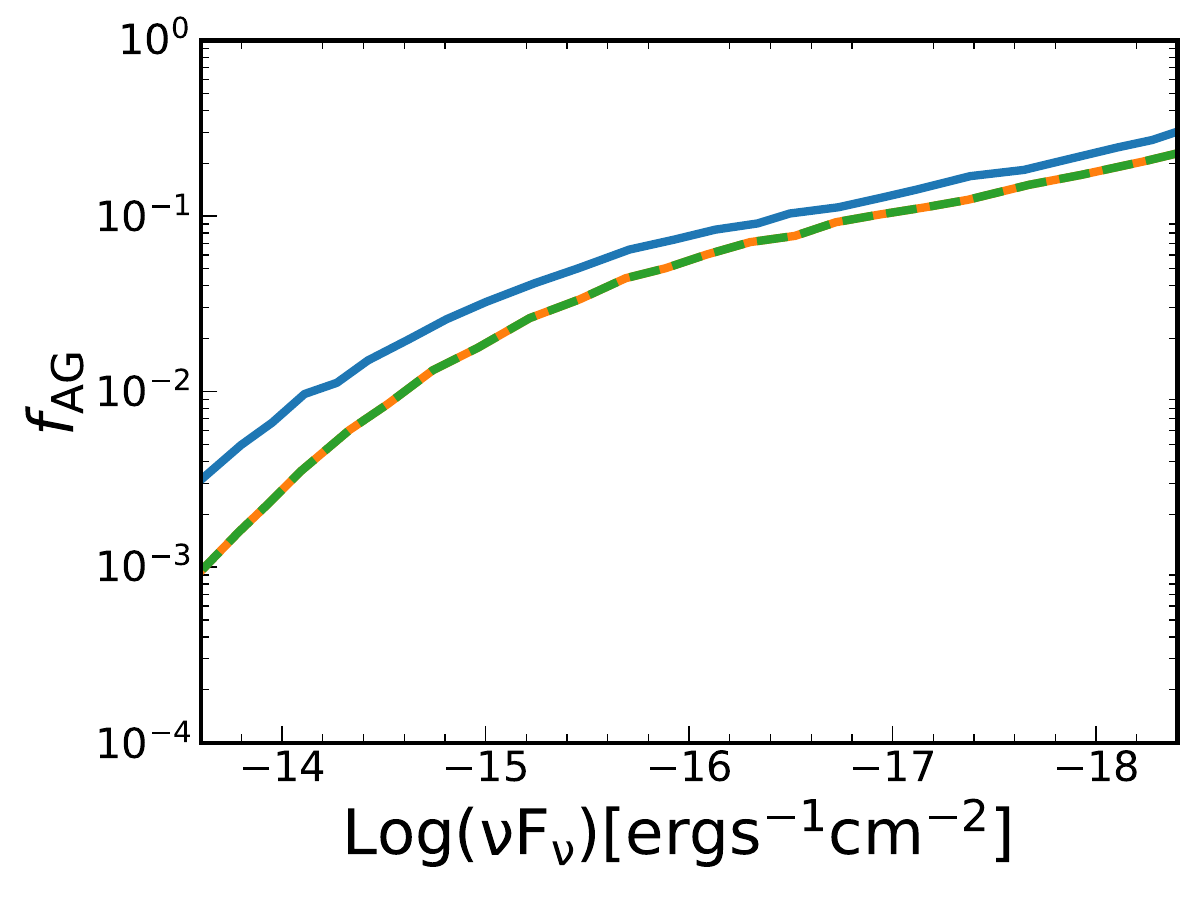}
\includegraphics[width=0.65\columnwidth]{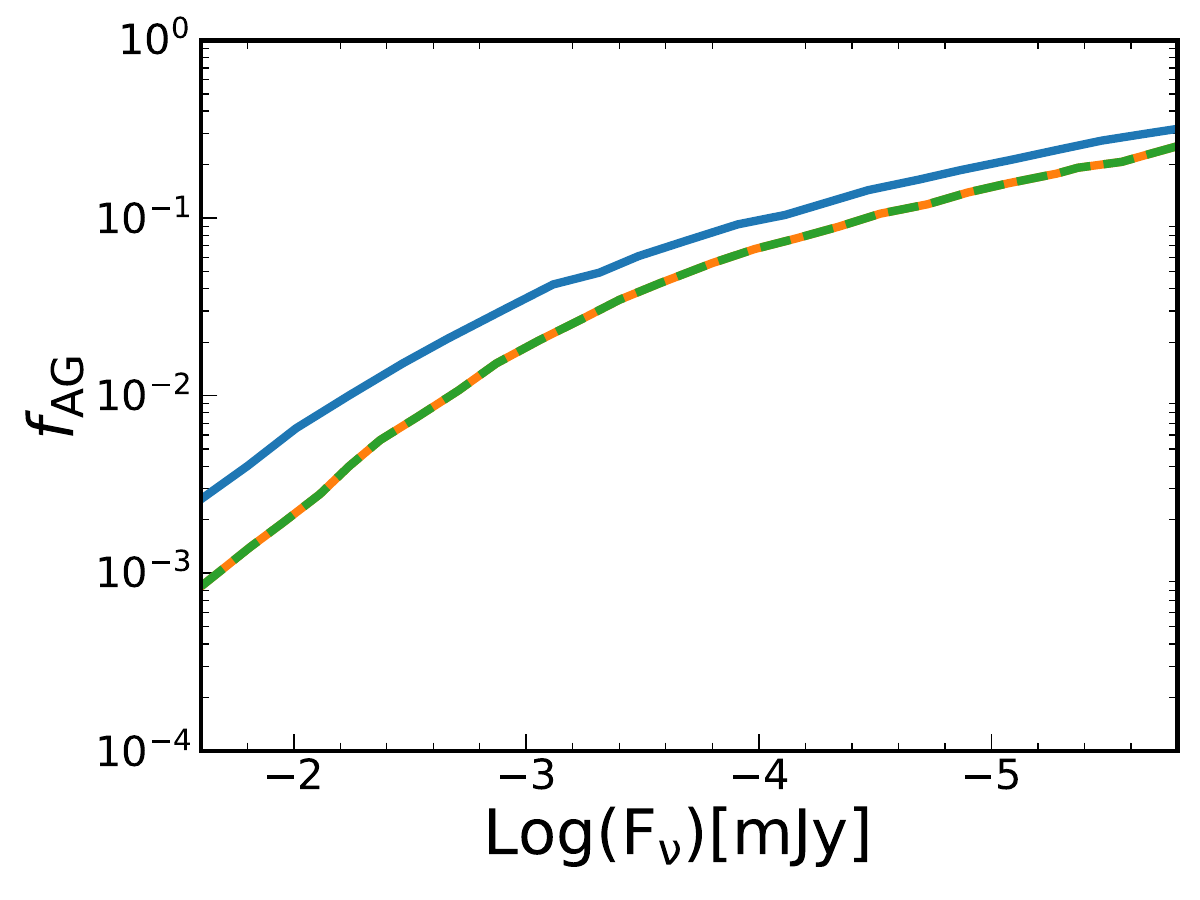}
\centering
\caption{
The dependence of the fraction of those lensed EM counterparts can be observed, i.e., $f_{\rm KN}$ and $f_{\rm AG}$ for kilonova (top) and afterglow (bottom) respectively.  In the top panels, the left, mid and right panel represent the results of $f_{\rm KN}$ for $\rm F106$, $\rm F158$, $\rm F213$ band of RST, while in the bottom panels, the left, mid and right panel represent $f_{\rm AG}$ for $\rm F158$, radio at $8.4$\,GHz and X-ray at $5\rm KeV$, respectively. In each panel, the blue solid, orange solid, and green dashed curves represent the results by using ET, CE, and the ET-CE network, respectively.
}
\label{fig:fem}
\end{figure*}

The factor related to the astrophysical radiation process is the fraction of {lensed kilonovae and afterglows} that can be observed by specific telescopes or detectors, denoted as $f_{\rm KN/AG}$ in Equation~\eqref{eq:flen}. We present the results of $f_{\rm KN}$ and $f_{\rm AG}$ with different detection thresholds in various wavelength bands in Figure~\ref{fig:fem}. In each panel, the blue solid, orange solid, and green dashed curves represent the results obtained by using ET, CE, and the ET-CE network, respectively.

The top left, middle, and right panels represent the results of $f_{\rm KN}$ for three different IR bands, i.e., $\rm F106$, $\rm F158$, $\rm F213$ of RST, respectively. {It is evident that the deeper the limiting magnitude, the higher the $f_{\rm KN}$, because deeper observations detect a larger fraction of lensed kilonovae.} Although the results for all the three bands are similar, differing at most by a factor of $1-2$, different GW detectors yield different $f_{\rm KN}$. For example, in the case of ET, the results are $\sim 5 $ times larger than those for CE and the ET-CE network. This is mainly because the lensed GW events detected by ET are at much lower redshifts, owing to ET's relatively lower sensitivity. We also note that this discrepancy decreases with increasing limiting magnitude, as almost all lensed kilonovae can be observed with sufficiently deep observations {($\gtrsim 32$\,mag)}. 

For the detectable lensed afterglow fraction, we also plot the results of $f_{\rm AG}$ in the bottom panels of Figure~\ref{fig:fem} for different wavelength bands: $\rm F158$ (left), radio at $8.4$\,GHz (middle), and X-ray at $5$\,KeV (right). Similarly to the results of $f_{\rm KN}$, the fraction $f_{\rm AG}$ increases with decreasing detection threshold. In addition, $f_{\rm AG}$ for ET is higher than that for CE and the ET-CE network, for the same reason discussed above. Despite these similarities, $f_{\rm AG}$ in the $\rm F158$ band is substantially smaller than $f_{\rm KN}$, which can be partly explained by the highly anisotropic radiation of the afterglow signals, i.e., only a certain fraction can be viewed from the proper angular direction. 

\subsection{Lensed GW + Kilonova/afterglow signals}

\begin{figure*}
\centering
\includegraphics[width=0.65\columnwidth]{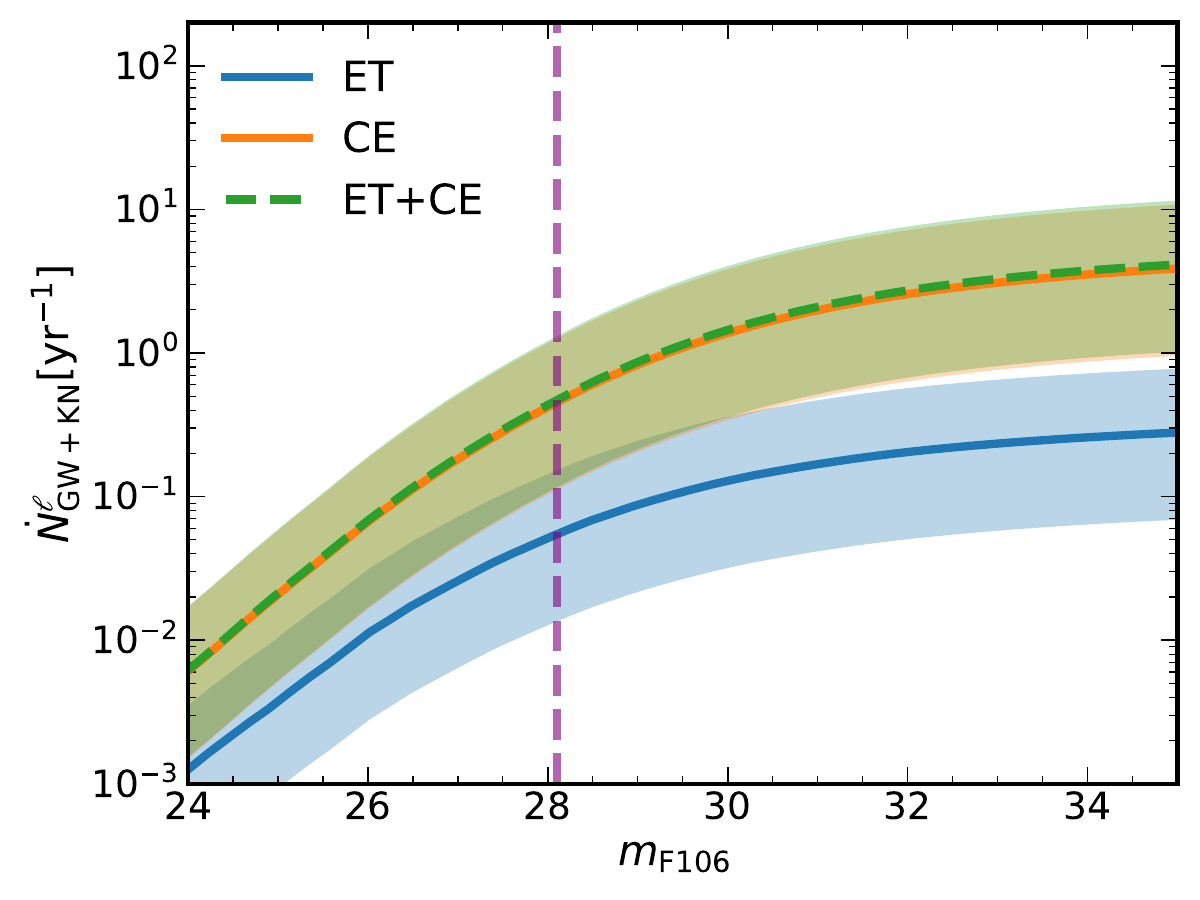}
\includegraphics[width=0.65\columnwidth]{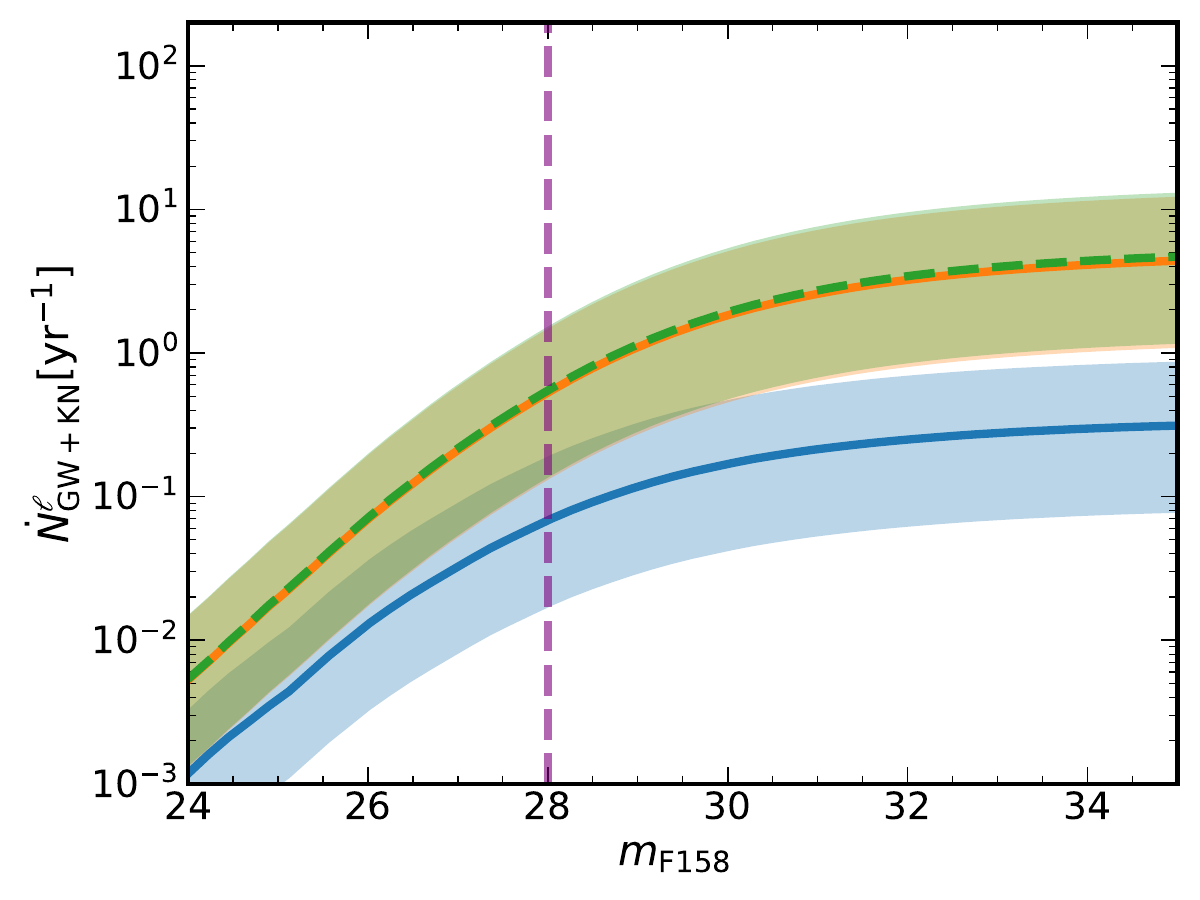}
\includegraphics[width=0.65\columnwidth]{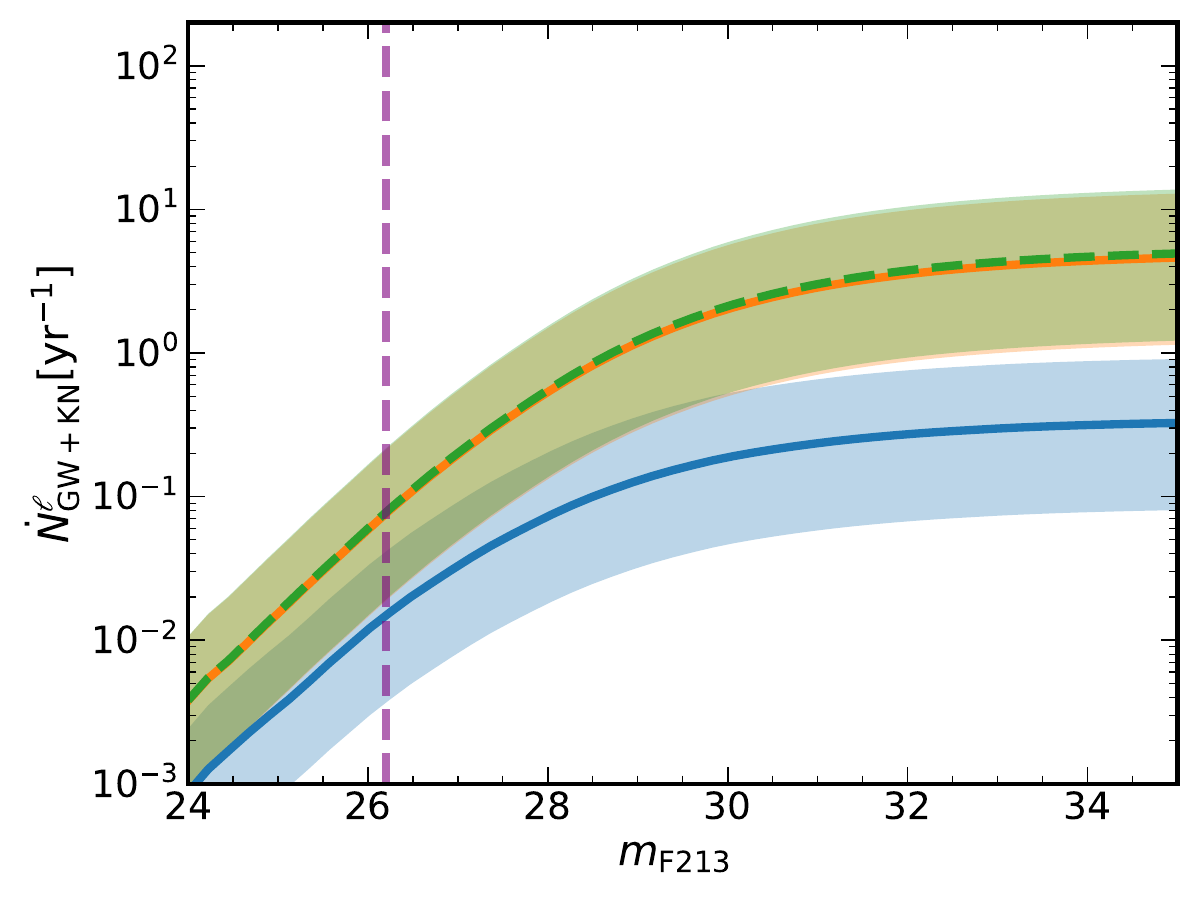}
\includegraphics[width=0.65\columnwidth]{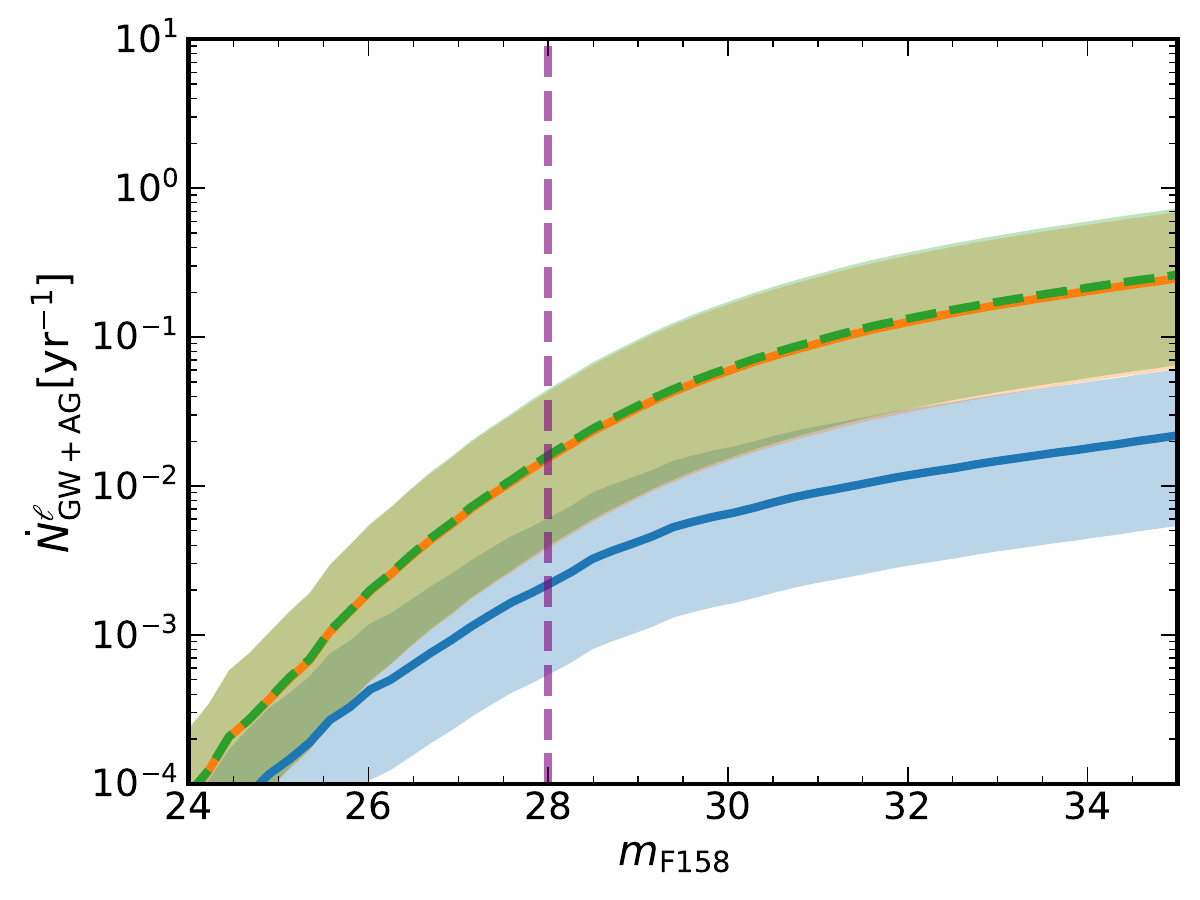}
\includegraphics[width=0.65\columnwidth]{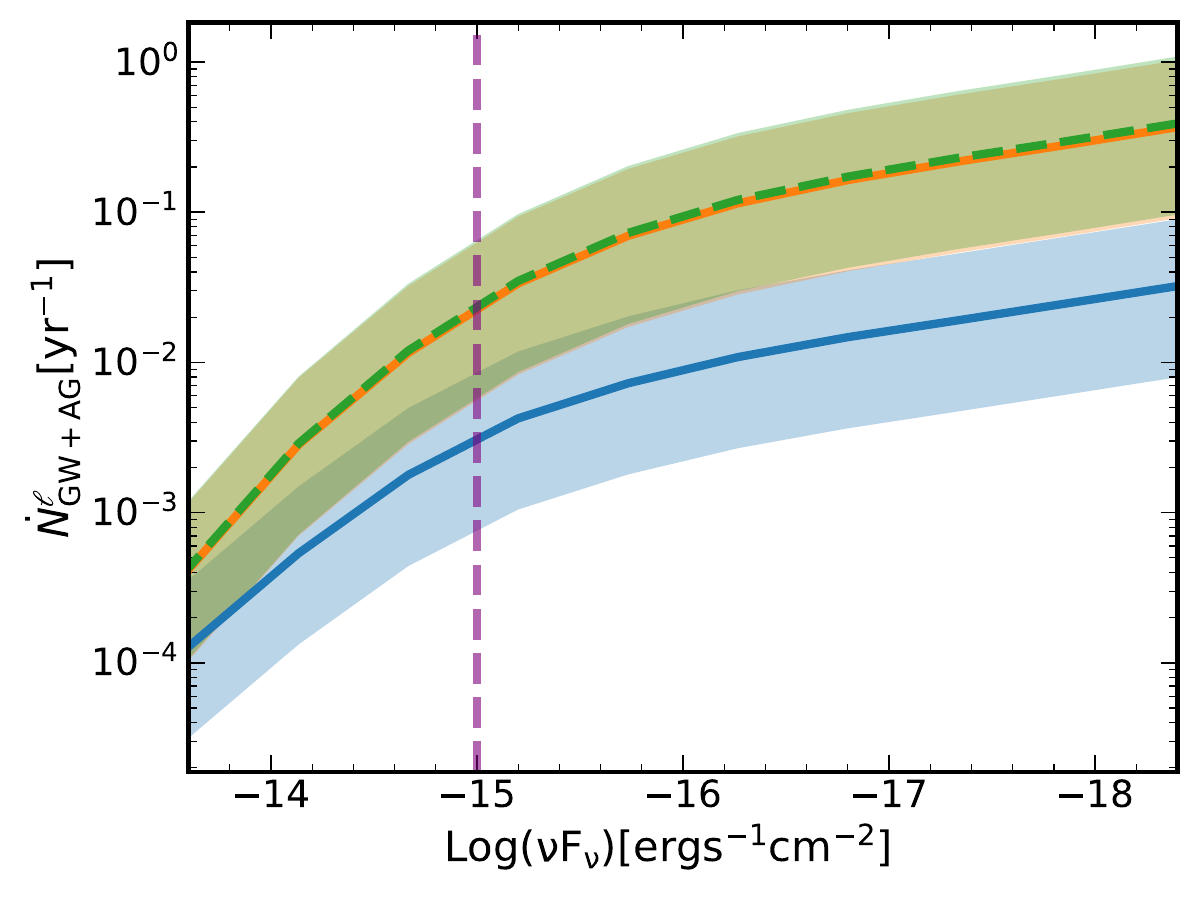}
\includegraphics[width=0.65\columnwidth]{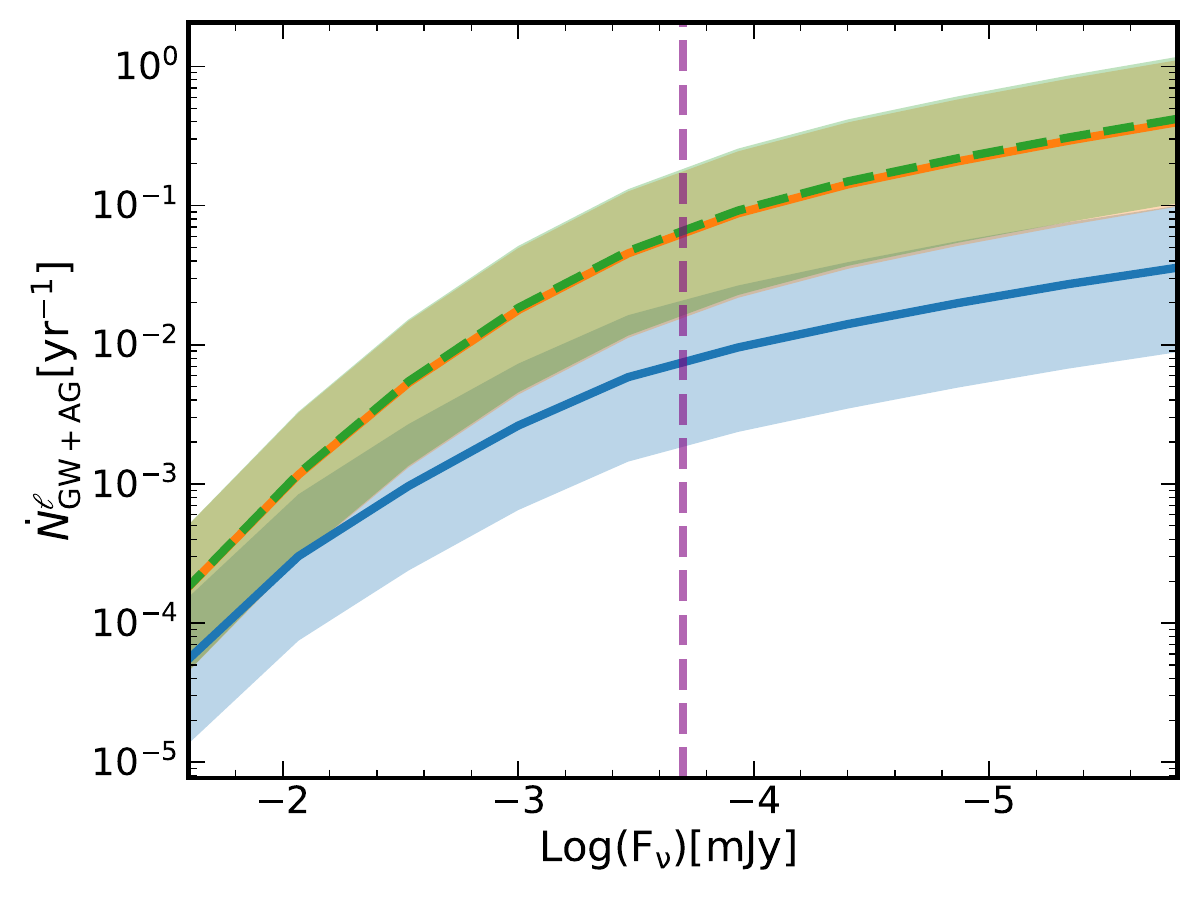}
\caption{
The {total} joint detection rate per year of the lensed GW associated with detectable lensed kilonova (top panels) and afterglow signals (bottom panels), i.e., $\dot{N}^{\ell}_{\rm GW+KN}$ and $\dot{N}^{\ell}_{\rm GW+AG}$.  In the top panels, the left, mid and right panel represent the results of  $\dot{N}^{\ell}_{\rm GW+KN}$ for $\rm F106$, $\rm F158$, $\rm F213$ band of RST, while in the bottom panels, the left, mid and right panel represent $\dot{N}^{\ell}_{\rm GW+AG}$ for $\rm F158$, radio at $\rm 8.4GHz$ and X-ray at $5\rm KeV$, respectively. In each panel, the blue solid, orange solid and green dashed represent the results by using ET, CE and their network. The shaded regions correspond to the error range induced by the estimation of local merger rate density of BNS mergers. The red purple dashed line show the detection threshold for RST (i.e., $28.1$, $28.0$, $26.2$ $\rm mag$ respectively), Chandra ($10^{-15} \rm ergs^{-1}cm^{-2}$) and SKA2 ($0.2\nu \rm Jy$). 
}
\label{fig:len_gw_k}
\end{figure*}

Using the lensed host fraction and the lensed EM fraction, we can estimate the detection rate of strongly lensed GW signals associated with kilonova and afterglow signals produced by BNS mergers, denoted as $\dot{N}^{\ell}_{\rm GW+KN}$ and $\dot{N}^{\ell}_{\rm GW+AG}$ respectively. 
{
Similar with lensed sGRBs, we find that double image cases dominate the lensed kilonova and afterglow signals. For example, as shown in Table~\ref{table:rate}, the joint detection rate of lensed kilonova signals with double, triple and quadruple cases by ET-CE network and {F106 band of RST-like instrument are about $\sim 0.35^{+0.63}_{-0.27}, 0.050^{+0.089}_{-0.038}$ and $ 0.047^{+0.084}_{-0.036}\rm yr^{-1}$, respectively.} Hence, we focus on the total detection rate in the following discussions. 
}

The top panels of Figure~\ref{fig:len_gw_k} show the dependence of the total $\dot{N}^{\ell}_{\rm GW+KN}$ on the limiting magnitude of different RST filters: F106 (left), F158 (mid), and F213 (right). The detection threshold for GW is assumed to be $\varrho_{\rm GW,0}=8$. From this figure, we can draw the following conclusions. First, the deeper the limiting magnitude, the higher the detection rate. For example, as shown in the first panel, one may detect {approximately $0.448/1.50/2.84 \rm yr^{-1}$ lensed events}, assuming $m_{\rm F106}$ to be $28/30/32$\,mag and the ET-CE network. In addition, the slope decreases with increasing $m_{\rm lim}$, indicating that the detection rate is rather limited by the EM telescope's performance than by the GW detector. Second, the detection rate at the same limiting magnitude for the CE-ET network is similar to that obtained for CE alone, but higher than that obtained for ET alone by a factor of $\sim 1-5$. This can be attributed to the higher sensitivity of CE, which dominates the performance of the ET-CE network. Third, there is no significant difference in the resulting detection rate when assuming the same GW detector but different bandwidths at the same limiting magnitude. {For example, the required limiting magnitudes to observe at least $1$\,yr$^{-1}$ such event are $\sim 29.2^{+5.51}_{-1.50}$, $28.8_{-1.27}^{+4.35}$, and $28.7^{+3.77}_{-1.16}$\,mag, } assuming the ET-CE network in these three bands, respectively. For the current design of RST, the limiting magnitude $m_{\rm F213}$ is much lower than the other two bands, i.e., $26.2 $\,mag. The resulting $\dot{N}^{\ell}_{\rm GW+KN}$ is {approximately $\sim 0.45/0.55/0.078$\,yr$^{-1}$ for ET-CE }network and F106/158/213 band of RST, indicating that with about ten years of accumulation, about $0.2-15$ such events can be detected. 

For the detection rate of lensed GW signals associated with afterglow signals, the results of $\dot{N}^{\ell}_{\rm GW+AG}$ based on the detection threshold are plotted in the bottom panels of Figure~\ref{fig:len_gw_k}. Compared to the GW+KN case, adopting the same limiting magnitude in the band of $\rm F158$, the detection rate is much lower. For example, assuming $\rm m_{\rm F158}\sim 28.1$\,mag, the joint detection rate $\dot{N}^{\ell}_{\rm GW+AG}$ for ET-CE {is about $0.016_{-0.012}^{+0.028}$\,yr$^{-1}$}, approximately four times smaller than $\dot{N}^{\ell}_{\rm GW+KN}$. This discrepancy can be explained by two main factors. First, the lensed fraction $f_{\rm AG}$ is much smaller than $f_{\rm KN}$ due to the viewing angle selection effect caused by the highly anisotropic radiation of the afterglow signals. Second, it is assumed that only when the post-merge remnant of the BNS merger is a HMNS/BH can the jet be produced by the Blandford-Znajek mechanism. Moreover, during the propagation of the jet through mass debris, the jet energy $E_{\rm jet}$ must exceed the breakout energy to successfully break out of the ejecta and interact with the interstellar medium to produce afterglow signals. Considering these criteria (see Appendix for more detailed descriptions), only $\sim 39.4\%$ BNS mergers in our mock sample can produce afterglow signals. Therefore, it is challenging to detect optical lensed afterglow signals. When it comes to radio afterglow detection, assuming the threshold of $0.2$\,$\mu$Jy, the joint detection rate {is about $0.07_{-0.05}^{+0.12}$\,yr$^{-1}$} for the network case. Therefore, we conclude that the joint detection of lensed GW + radio afterglow may be more optimistic than in the optical band, taking advantage of the high sensitivity of SKA2.

{For X-ray afterglow detection, high-energy missions such as Swift \citep{2005SSRv..120..165B}, SVOM \citep{2022IJMPD..3130008A}, and Einstein Probe \citep{2022hxga.book...86Y} can provide prompt or early detections and localizations for bright or highly magnified events, and may trigger deeper follow-up observations \citep{2025RSPTA.38340118R}. Since their sensitivities are generally shallower than those required for most faint lensed afterglows considered here, we use Chandra-like and ATHENA-like thresholds as representative benchmarks for pointed X-ray follow-up.} Assuming the Chandra detection threshold, we find $\dot{N}^{\ell}_{\rm GW+AG}$ is {about $\sim 0.003$, $0.024$, and $0.024$\,yr$^{-1}$ for} ET, CE, and the ET-CE network, respectively, which is still rare. The upcoming next-generation X-ray telescope, such as ATHENA, is much more powerful, with a detection threshold of about $2\times10^{-17}$\,erg\,s$^{-1}$\,cm$^{-2}$ \citep{2012arXiv1207.2745B}. With ATHENA and ET-CE network, for example, {the joint detection rate is $0.16_{-0.12}^{+0.29}$\,yr$^{-1}$}, indicating that with ten years of accumulation, it is possible to detect $\sim 0.5-5$ lensed GW events associated with lensed X-ray afterglow signals.

\subsection{Comparison with MLG+23}

In this section, we compare our results on the joint detection rate of lensed GW signals associated with kilonova and afterglow signals produced by BNS mergers ($\dot{N}^{\ell}_{\rm GW+KN}$ and $\dot{N}^{\ell}_{\rm GW+AG}$, respectively) with the results from MLG+23. Note that our strategy corresponds to the ``conservative criterion" in MLG+23, where the faintest image of both the GW and EM signals has an S/N or flux above the detection threshold. Moreover, the results in MLG+23 are given under the Vega magnitude system. Therefore, we convert the Vega magnitudes to AB magnitudes using the zero-point flux of Vega, i.e., 
\begin{equation}
m_{\rm F158}^{\rm AB}=m_{\rm F158}^{\rm Vega}+2.5\rm log\left(\frac{3631\rm Jy}{1090\rm Jy}\right),
\end{equation}
where $1090$\,Jy is the flux of Vega in the F158-band. 

We first compare our results on $\dot{N}^{\ell}_{\rm GW+KN}$ with those obtained by MLG+23 (see Figure 5 therein). Assuming the same $f_{\rm Host}$ and median local merger rate density, the joint detection rate is about $\sim 0.83$\,yr$^{-1}$ for the {RST-like} F158 band, which is slightly higher than the results presented in this paper. Note that MLG+23 adopts the isotropic two-component kilonova model from \citet{2017ApJ...851L..21V}, with intrinsic parameters chosen to match those of GW170817, which may be an oversimplification.

Regarding the joint detection rate of lensed GW and afterglow signals $\dot{N}^{\ell}_{\rm GW+AG}$ in the {RST-like}/F158-band, the result in MLG+23 is $0.028\rm yr^{-1}$, which is about two times higher than the results presented in this paper, despite the significantly different methodologies. The difference can be explained as follows. First, as previously discussed, only a fraction of $\sim 39.4\%$ BNS mergers can produce observable afterglow signals due to successfully breakout jet structures, a factor ignored in MLG+23. Second, the model parameters of the afterglow LCs in MLG+23 were chosen to be the same as those obtained in \citet{2017ApJ...851L..21V} for injection into {Afterglowpy}, which results in systematically fainter LCs. Third, we adopt the EPL profile as the lensing model, which will produce systematically lower magnification factors compared with the SIE lensing model adopted in MLG+23. By the above reasons, we obtain slightly smaller results than that in MLG+23. 

\section{Conclusions and Discussions}
\label{sec:con}

In this paper, we utilize the results of binary population synthesis, numerical simulation of BNS mergers, jet propagation models, and observational constraints from GW170817, to explore the joint detection rate of strongly gravitationally lensed multi-messenger signals from BNS mergers. Our main conclusions are summarized as follows.

\begin{itemize}
\item The sGRB signals associated with the lensed GW signals are challenging to detect with the current sensitivity of the Fermi-GBM. However, if the sensitivity of the $\gamma$-ray detector is enhanced by a factor of $5$ compared to Fermi-GBM (e.g., $<3\times 10^{-8}$\,erg\,cm$^{-2}$), it is possible to detect a lensed GRB event within a 30-year observation period.
%

\item {Large GW localization uncertainties and the faintness of lensed kilonovae and afterglow motivate a complementary pointed follow-up strategy. In this strategy, pre-identified galaxy-scale lens candidates within the GW localization region are targeted with deep observations, which may improve the per-lens sensitivity when suitable lens catalogs and telescope resources are available.}

\item Adopting such a strategy, the fraction of host galaxies that can be identified as lensed systems, $f_{\rm Host}$, is of great significance. By simulating the host galaxies, we find that this fraction is strongly dependent on the filter and limiting magnitude of the survey telescope(s) used to construct the lensing galaxy catalog. For the fiducial deep lens-catalog case considered here, we find
$f_{\rm Host}\simeq0.15$ and $0.3$ for ET and CE, respectively.

\item The joint detection of lensed GW associated with kilonova signals is promising. {We find that the detection rate, $\dot{N}^{\ell}_{\rm GW+KN}$, is approximately $0.45_{-0.34}^{+0.81}/0.55_{-0.41}^{+0.98}/0.078^{+0.139}_{-0.059}$\,yr$^{-1}$} with the current design of F106/158/213 bands of RST. 
\item The joint detection of lensed GW associated with afterglow signals is challenging to achieve in the radio and optical bands, even with the capabilities of SKA and RST. However, utilizing a future powerful X-ray telescope like ATHENA, we may detect approximately $0.5-5$ lensed GW events associated with lensed X-ray afterglow signals over a ten-year observational period. 
\end{itemize}

Note that we have made several simple assumptions and approximations in this work, which may affect the results. In reality, there are many complexities that need to be considered for a more robust investigation.  
{
First, we utilize the compact binary population synthesis model $\boldsymbol{\alpha10.\rm kb\beta0.9}$ for generating the mock BNS merger population.
Though this model was demonstrated to be most compatible with both current observations on both the Galactic and GW BNS systems \citep{2022MNRAS.509.1557C}, one may still expect substantial uncertainties on 
the merger rate density evolution for the extremely complex physical processes involving in the evolution of binaries. Nevertheless, the accumulation of BNS mergers will provide better measurements on the BNS merger rate and provide more strong constraints on the population synthesis model, thus lead to a more robust estimation on the joint lensed rates of multi-messenger signals. 
}

Second, in estimating lensed kilonova signals, we do not consider the possible energy injection by the long-lived supramassive neutron star (SMNS) merger remnant (i.e., magnetar). The additional energy injection from magnetic wind could enhance the kilonova by several times \citep[e.g.,][]{2024MNRAS.527.5166W, chen24}, potentially leading to a more optimistic detection rate. As discussed in \citet{chen24}, the detection rate $\dot{N}^{\ell}_{\rm GW+KN}$ may be enhanced by a factor of $2$ or more. 

Third, when estimating the lensed afterglow, we assume that only the hypermassive neutron star (HMNS)/black hole (BH) remnant can launch the jet via the Blandford-Znajek effect. However, recent numerical simulations \citep[e.g.,][]{2024arXiv240503705B} have shown that a SMNS remnant can also form a jet-like structure, with energy approximately $10$ times larger than that of the Blandford-Znajek jet. Moreover, in this paper, we assume the conversion efficiency of jet energy into radiation, $\eta_{\gamma}$, to be uniformly distributed between $0.1$ and $0.3$, similar to the results from GW170817. However, this efficiency, $\eta_{\gamma}$, differs significantly from the constraints derived from observed sGRBs with afterglow signals, assuming that all sGRBs originate from BNS merger. As discussed in \citet{2015ApJ...815..102F}, $\eta_{\gamma}$ can vary widely between $6.6\times10^{-3}$ and $0.97$, with medians of $0.56$ and $0.4$ for $\epsilon_{\rm B}=0.1$ and $0.01$, respectively. The uncertainty in $\eta_{\gamma}$ may lead to an uncertainty in our estimation of $\dot{N}^{\ell}_{\rm GW+AG}$ and $\dot{N}^{\ell}_{\rm GW+sGRB}$ by a factor of approximately $5$ or more.

{
We also note that 
recently Vera C. Rubin Observatory (Rubin) \citep{2019ApJ...873..111I} releases its first light, marking the opening of a new era for time-domain astronomy. Owing to its large FOV and deep limiting magnitudes, it will be transformational for applications of almost all aspects in gravitational lensing \citep[][]{2025RSPTA.38340117S}. 
It is thus expected that Rubin will also play an important role in detection of lensed muti-messenger signals of BNS mergers. For example, \citet{2025RSPTA.38340118R} provided a follow-up strategy of EM counterparts to highly magnified lensed GW events with the help of wide FOV optical telescopes such as Rubin by covering as much of a GW sky localization area, which will be efficient and economic for LVK O4 and future runs. 
{Unlike the wide-field follow-up ToO strategy, we highlight the importance of the identification of the lensed host galaxies with next-generation GW detectors and then targeting those hosts with high spatial-resolution, space-borne telescopes to capture their lensed EM signals. We recognize that Target of Opportunity (ToO) observations for space-borne facilities are highly restricted due to their disruptive impact on scheduled programs and the stringent criteria of Time Allocation Committees (TACs) \citep[e.g.,][]{2024APh...15502904A}. Nevertheless, our strategy offers a {complementary} pathway to search for faint lensed sources, as it significantly reduces the required telescope time and number of pointings. Here we also note that similar strategies involving the lensed host galaxies have also been proposed for searching the stellar binary black hole mergers without EM counterparts \citep[e.g.,][]{2024MNRAS.530.3368W,2025RSPTA.38340152U}. While these studies emphasize that such searches face significant challenges in the current LVK era due to poor GW localization and catalog incompleteness, they also mentioned that these hurdles are likely to be overcome in the era of next-generation detectors like ET and CE due to their increased sensitivities \citep[see the discussion in Section  4 (a) in ][]{2025RSPTA.38340152U}. Specifically, as discussed in \citet{chen25}, the effective network may help to reduce localization uncertainties to $\lesssim 1 \rm ~deg^2$ (and even $\lesssim 0.1$\,deg$^2$ in a significant fraction of cases) and thus making our strategy a nice addition in the next generation GW detection new era.}

{
One may also note that lensing by galaxy clusters (or groups) is not considered due to its complexity in this paper. Such lensing can also
produce multiple images of BNS mergers associated with multi-messenger signals \citep[e.g.,][]{2018MNRAS.475.3823S, 2022arXiv220412977S}. 
For example, \citet{2024ApJ...977...64C} found that the the lensing by galaxy clusters may contribute 
to the total number of strongly lensed BNS multi-messenger signals similarly as that by the lensing of galaxies, primarily due to the larger lensing cross section and magnification factor of galaxy clusters. This could aid in detecting high-redshift BNS mergers and constraining the formation and evolution of BNSs in the early universe, despite the complexities of cluster (or group) lensing modeling due to multiple mass components and intricate caustics.  
}

\section*{acknowledgement}
This work is partly supported by 
the National Natural Science Foundation of China (Grant Nos. 12273050, 11991052), the Strategic Priority Program of the Chinese Academy of Sciences (Grant No. XDB XDB0550300), and the National Key Program for Science and Technology Research and Development (Grant Nos. 2020YFC2201400 and 2022YFC2205201).

\bibliographystyle{aasjournal}
\bibliography{ref.bib}

\appendix

\section{Merger scenario}  

\subsection{Ejecta mass}
\label{app:B1}
In this work, we consider three different types of ejecta, i.e., dynamical, viscous and wind ejecta. When a BNS merges, there is a small fraction of neutron-rich matter, normally $10^{-4}-10^{-2} M_{\odot}$ for BH remnant case and $\sim 10^{-1} M_{\odot}$ for NS remnant case, respectively, can be ejected out with a velocity of $\sim 0.1-0.3c$ within the dynamical timescale ($\sim \rm ms$) by the tidal forces or by shocks in the collision of the neutron star cores \citep[e.g., ][]{2017LRR....20....3M}. Numerical simulations \citep[e.g., ][]{2020PhRvD.101j3002K} have shown that the mass of the dynamical ejecta in the BH case can be approximated by 
\begin{equation}
\frac{m_{\mathrm{dyn}}}{10^{-3} M_{\odot}}=\left(\frac{a}{C_1}+b \frac{m_2^n}{m_1^n}+c C_1\right) m_1+(1 \leftrightarrow 2),
\label{mass:dyn}
\end{equation}
where $C_{1,2}$ is the compactness of the NS and the fitted parameters are $a=-9.3335$, $b=114.17$, $c=-337.56$, and $n=1.5465$. While the norm $v_{\rm dyn}$ and opening angle $\theta_{\rm dyn}$ of the ejecta velocity are expressed as \citep{2017CQGra..34j5014D}: 
\begin{equation}
\begin{gathered}
\theta_{\mathrm{dyn}} \simeq \frac{-2^{\frac{4}{3}} v_\rho^2+2^{\frac{2}{3}}\left[v_\rho^2\left(3 v_z+\sqrt{9 v_z^2+4 v_\rho^2}\right)\right]^{\frac{2}{3}}}{\left[v_\rho^5\left(3 v_z+\sqrt{9 v_z^2+4 v_\rho^2}\right)\right]^{\frac{1}{3}}}, \\
v_{\mathrm{dyn}} \simeq\left[f_1\left(1+f_3 C_1\right) \frac{m_1}{m_2}+\frac{f_2}{2}\right]+(1 \leftrightarrow 2),
\label{theta:dyn}
\end{gathered}
\end{equation}
where $f_1=-0.3090$, $f_2=0.657$ $f_3=-1.879$, $v_{\rho}$, and $v_z$ are the ejecta velocity component in the cylindral coordinate given by \citet{2017CQGra..34j5014D}.

On the other hand, a fraction of NS decompressed matter may be centrifugally supported and therefore produce an accretion disk around the merger remnant, which may also contribute significantly to the ejecta mass through outflows motivated by the neutrino wind near the symmetric axis \citep[e.g., ][]{2017LRR....20....3M}. The mass of this wind ejecta, $m_{\rm wind}$ is often linked with the mass of the remnant disk $m_{\rm disk}$ by $\xi_{w}$, i.e., $m_{\rm wind}=\xi_{w}m_{\rm disk}$ \citep[e.g.,][]{2022ApJ...937...79C, 2023MNRAS.522..912Z}. In this work, we evaluate $m_{\rm disk}$ by
the results of numerical simulation \citep[e.g., ][]{2020PhRvD.101j3002K} as
\begin{equation}
m_{\rm disk}=m_1 \rm {\max}\left(5\times 10^{-4}, (aC_{1}+c)^d\right),
\label{mass:disk}
\end{equation}
where $a=-8.1324$, $c=1.4820$, and $d=1.7784$. 

The viscous torques of the accretion disks around the massive NSs or BH can also unbind ejecta matters, which may be much more massive than the wind ejecta driven by neutrino wind on viscous time scale. As in wind ejecta, we link the mass of the viscous ejecta $m_{\rm vis}$ with the mass of the remnant disk $m_{\rm disk}$ by $\xi_{v}$: $m_{\rm vis}=\xi_{v}m_{\rm disk}$ \citep[e.g.,][hereafter C22]{2022ApJ...937...79C}.

\subsection{Jet Energy}
\label{app:jet}

In this section, we basically follow the procedure in C22 to estimate the energy of the relativistc jet and consider its breakout. More detailed information can be seen in Appendix B.3 of C22.  One possible supply for the jet energy is the spinning BH with a magnetized accretion disk, i.e., the Blandford-Znajek (BZ) mechanism. In this scenario, the total injected energy of the jet can be written as 
\begin{equation}
E_{\rm jet,0}=\epsilon_{\rm BZ}(1-\xi_{\rm w}-\xi_{\rm s})m_{\rm disk}c^2\Omega_{\rm H}^2f(\Omega_{\rm H}),
\end{equation}
where $\epsilon_{\rm BZ}=0.022$ is the dimensionless factor, $m_{\rm disk}$ is the mass of the accretion disk and $\Omega_{\rm H}=a_{\rm H}/2(1+\sqrt{1-a_{\rm H}^2})$ and $f(\Omega_{\rm H})=1+1.38\Omega_{\rm H}^2-9.2\Omega_{\rm H}^4$ with $a_{\rm H}$ represents the dimensionless spin. The mass of the remnant $m_{\rm rem}$ can be estimated by the energy conservation of BNS before and after the merger as
\begin{equation}
m_1+m_2=m_{\rm rem}+m_{\rm GW}+m_{\rm disk}+m_{\rm dyn},
\label{eq:mass}
\end{equation}
where $m_{\rm GW}$ is the energy dissipation by GW radiation, which can be calculated using the fitting formula considering the inspiral to the post-merger stage \citep[e.g.,][]{Bernuzzi:2014kca, 2018PhRvL.120k1101Z}. To fulfill the prior condition for the BZ mechanism, it is necessary to identify the remnant object to be a rapidly rotating BH (or short-lived HMNS) by its mass, i.e, if $\rm m_{\rm rem}>1.2 M_{\rm TOV}$. 

After launching, the jet propagates in the mass ejecta, the sGRB and afterglow signals can be produced only when the jet can successfully break out. Following \citet{2018ApJ...866....3D}, the breakout energy is set as
\begin{equation}
E_{\rm bkt}=0.05\theta_{\rm j,0}E_{\rm ej},
\end{equation}
where $\theta_{\rm j,0}$ is the initial opening angle of the jet, assuming to be $15^{\circ}$, and $E_{\rm ej}$ is the energy of the ejecta, which can be calculated by the mass profiles of the ejecta. Then the available jet energy can be evaluated for the sGRB prompt emission and the kinetic energy of afterglow as 
\begin{equation}
E_{\rm tot,\gamma}=\eta_{\gamma}(E_{\rm jet,0}-E_{\rm bkt}),
\label{gamma}
\end{equation}
and 
%
\begin{equation}
E_{\rm tot,K}=(1-\eta_{\gamma})(E_{\rm jet,0}-E_{\rm bkt}),
\label{kinetic}
\end{equation}
respectively, where $\eta_{\gamma}$ is the conversion efficiency of jet energy into radiation. Note that the value of $\eta_{\gamma}$ is still under debate due to the uncertainty in the afterglow fitting procedure. In this paper, we set $\eta_{\gamma}$ to be uniformly distributed in $[0.1,0.3]$ for illustration purposes. With Equations~\eqref{gamma} and \eqref{kinetic}, we can estimate the fluence of sGRB prompt emission and the afterglow fluxes, by assuming a Gaussian-structured jet. Assuming the masses of the two components of GW170817 to be $m_1=1.46 M_{\odot}$ and $m_2=1.27 M_{\odot}$, we can estimate the kinetic energy of the GW170817 afterglow to be $E_{\rm tot,K}\sim 3.5\times 10^{49} ~\rm erg$ using the above Equations. 

\section{Fitting Data to Model}
\label{app:Fit}

In this appendix, we show the detailed procedures of the Bayesian fitting of the afterglow of GW170817. All data points are drawn from the full uniform dataset presented in \citet{Makhathini:2020ece} and then transferred into three bands, i.e., $5~\rm keV$, $590~\rm nm$ and $6~\rm GHz$ as done in \citet{Makhathini:2020ece}. We adopt the Python Package {AfterglowPy} for the Gaussian structured Jet model \citep{2020ApJ...896..166R} to fit the data points, after fixing the isotropic equivalent energy to be $E_{0}=7\times 10^{51} ~\rm erg$. This choice of $E_0$ is the direct result from the Gaussian structured jet, adopting the value of the kinetic energy $E_{\rm tot,K}\sim 3.5\times 10^{49} \rm erg$ and $t_{\rm c}$ to be $0.07~\rm rad$. Using the well-documented Monte-Carlo Sampling package {emcee} \citep{ForemanMackey2012emceeTM}, we generate the posterior distribution of 5 intrinsic parameters, i.e., $t_{\rm v}$, $n_{e,0}$, $p$, $\epsilon_{e}$ and $\epsilon_{e,B}$ utilizing 300 random walkers to run at least 5000 steps. The uniform prior and the fitted median value with $16,84\%$ error of these parameters are listed in Table~\ref{table:para}. Figure~\ref{fig:170817} shows the multiband LCs of the afterglows of GW170817 and the corresponding best model fit. 

\begin{table}[h]
\caption{Model parameters of the Gaussian Structured Jet model and their fitted values of afterglow of GW170817}
\label{table:para} 
\centering
\begin{tabular}{lcccc} 
        Parameter  & Name & Prior range & Posterior value  \\

\hline    \hline
       $t_{\rm v}$ & Viewing angle &  $\rm U[0,0.8]$ & $0.41_{-0.01}^{+0.01}$  \\

       ${\rm log}(n_{e,0})$ &  Number density of electrons  in ISM  & $\rm U[-6,1]$  & $-3.85_{-0.09}^{+0.08}$   \\
       $p$ & Power law index of accelerated shock & $\rm U[2,5]$  & $2.12_{-0.01}^{+0.01}$   \\

       ${\rm log}\epsilon_{e}$ & accelerated electron energy fraction & $\rm U[-5,-0.5]$ & $-1.09_{-0.03}^{+0.04}$    \\
       ${\rm log}\epsilon_{e,B} $ &  magnetic field energy fraction & $\rm U[-5,-0.5]$ & $-1.82_{-0.09}^{+0.08}$ \\
       \hline \hline
\end{tabular}
\end{table}

\begin{figure}
\includegraphics[width=0.8\columnwidth]{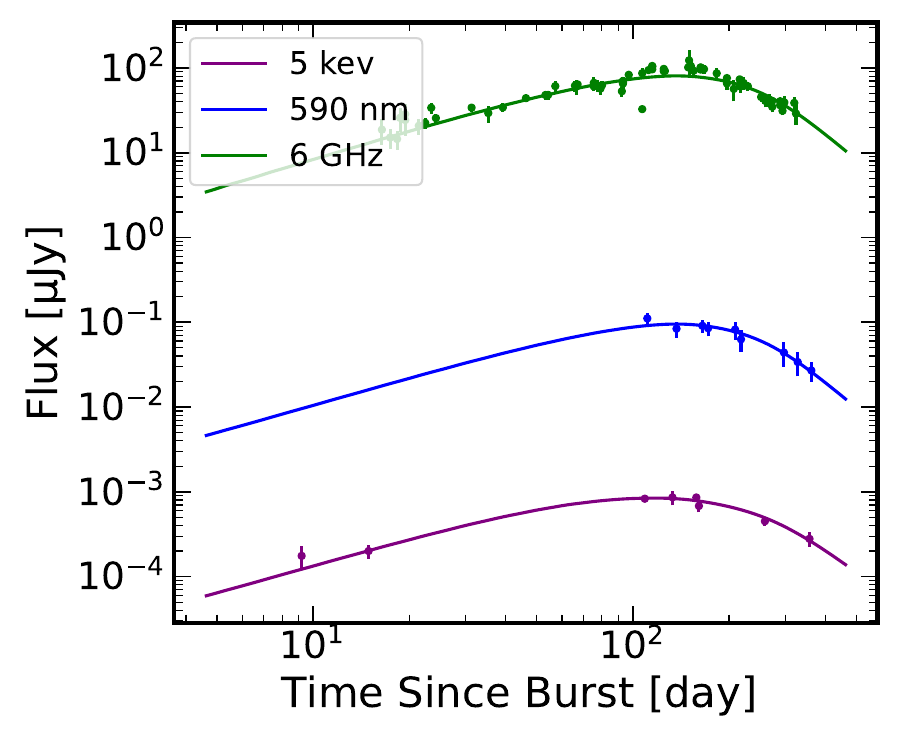}
\centering
\caption{Multiband light curves of afterglow of GW170817 and the best model fit. The green, blue and purple symbols with error bar represent the data points transferred into X-ray ($5\rm ~keV$), optical ($590~\rm nm$) and radio ($\rm 6~GHz$) band. The corresponding solid lines show the best-fit results.
}
\label{fig:170817}
\end{figure}

\end{document}